\documentclass[12pt,a4paper]{article}
\usepackage{amsmath,amssymb,psfrag,graphicx,axodraw}
\usepackage{hyperref}
\begin{document}
\input{epsf}
\topmargin 0pt
\oddsidemargin 5mm
\headheight 0pt
\headsep 0pt
\topskip 9mm
\voffset 1cm

\pagestyle{empty}

\newcommand{\beq}{\begin{equation}}
\newcommand{\eeq}{\end{equation}}
\newcommand{\bea}{\begin{eqnarray}}
\newcommand{\eea}{\end{eqnarray}}
\newcommand{\rf}[1]{(\ref{#1})}
\newcommand{\pa}{\partial}
\newcommand{\nn}{\nonumber}

\newcommand{\e}{\mbox{e}}
\renewcommand{\d}{\mbox{d}}
\newcommand{\g}{\gamma}
\renewcommand{\l}{\lambda}
\renewcommand{\L}{\Lambda}
\renewcommand{\b}{\beta}
\renewcommand{\a}{\alpha}
\newcommand{\n}{\nu}
\newcommand{\m}{\mu}
\newcommand{\Tr}{\mbox{Tr}}
\newcommand{\E}{\mbox{E(q)}}
\newcommand{\Ee}{\mbox{E}}
\newcommand{\K}{\mbox{K(q)}}
\newcommand{\Kk}{\mbox{K}}
\newcommand{\ep}{\varepsilon}
\newcommand{\om}{\omega}
\newcommand{\del}{\delta}
\newcommand{\Del}{\Delta}
\newcommand{\sg}{\sigma}
\newcommand{\vph}{\varphi}
\newcommand{\sn}{\mbox{sn}}
\newcommand{\dn}{\mbox{dn}}
\newcommand{\cn}{\mbox{cn}}
\newcommand{\CA}{\mathcal{A}}
\newcommand{\oh}{\frac{1}{2}}
\newcommand{\oq}{\frac{1}{4}}
\newcommand{\dg}{\dagger}
\newcommand{\R}{\mathbb{R}}
\newcommand{\tr}{\mbox{Tr}\;}
\newcommand{\ra}{\right\rangle}
\newcommand{\la}{\left\langle}
\newcommand{\prt}{\partial}
\newcommand{\mi}{\!-\!}
\newcommand{\equ}{\!=\!}
\newcommand{\pl}{\!+\!}
\newcommand{\CN}{\mathcal{N}}
\newcommand{\cD}{{\cal D}}
\newcommand{\cS}{{\cal S}}
\newcommand{\cM}{{\cal M}}
\newcommand{\cK}{{\cal K}}
\newcommand{\cT}{{\cal T}}
\newcommand{\cN}{{\cal N}}
\newcommand{\cL}{{\cal L}}
\newcommand{\cO}{{\cal O}}
\newcommand{\cR}{{\cal R}}
\newcommand{\CH}{\mathcal{H}}
\newcommand{\CL}{\mathcal{L}}
\newcommand{\CO}{\mathcal{O}}
\newcommand{\CI}{\mathcal{I}}
\newcommand{\CT}{\mathcal{T}}
\newcommand{\CS}{\mathcal{S}}
\newcommand{\CM}{\mathcal{M}}
\newcommand{\CQ}{\mathcal{Q}}
\newcommand{\CE}{\mathcal{E}}
\newcommand{\CB}{\mathcal{B}}
\newcommand{\tF}{{\tilde{F}}}
\newcommand{\tL}{{\tilde{\L}}}
\newcommand{\tX}{{\tilde{X}}}
\newcommand{\tY}{{\tilde{Y}}}
\newcommand{\tZ}{{\tilde{Z}}}
\newcommand{\ty}{{\tilde{y}}}
\newcommand{\tz}{{\tilde{z}}}
\newcommand{\tg}{{\tilde{g}}}
\newcommand{\tG}{{\tilde{G}}}
\newcommand{\tH}{{\tilde{H}}}
\newcommand{\tT}{{\tilde{T}}}

\newcommand{\SL}{{\sqrt{\L}}}
\newcommand{\tSL}{\sqrt{\tL}}
\newcommand{\FL}{\L^{1/4}}
\newcommand{\bZ}{{\bar{Z}}}
\newcommand{\bX}{{\bar{X}}}

\newcommand{\remark}[1]{{\renewcommand{\bfdefault}{b}\textbf{\mathversion{bold}#1}}}

\vspace*{100pt}

\begin{center}

{\large \bf {Non-planar ABJM Theory and Integrability}}

\vspace*{26pt}

{\sl Charlotte Kristjansen}, {\sl Marta Orselli} and
{\sl Konstantinos Zoubos}

\vspace{10pt}
\vspace{10pt}

The Niels Bohr Institute, Copenhagen University\\
Blegdamsvej 17, DK-2100 Copenhagen \O , Denmark.\\ \vspace{.4cm}
{\tt\small kristjan@nbi.dk, orselli@nbi.dk, kzoubos@nbi.dk} 
\vspace{14pt}



\end{center}

\begin{abstract}
\noindent
Using an effective vertex method we explicitly derive the two-loop dilatation 
generator  of ABJM theory in its $SU(2)\times SU(2)$ sector,
including all non-planar corrections. Subsequently, we apply this generator 
to a series of finite length operators as well as to two different types of BMN
operators. As in ${\cal N}=4$ SYM,
at the planar level the finite length operators are found to exhibit a
degeneracy between certain pairs of operators with opposite parity -- a
degeneracy which can be attributed to the existence of an extra conserved
charge and thus to the integrability of the planar theory. When non-planar
corrections are taken into account the degeneracies between parity
pairs disappear hinting the absence of higher conserved charges.
The analysis of the BMN operators resembles
that of ${\cal N}=4$ SYM. Additional non-planar terms appear
for BMN operators of finite length but once the strict BMN limit is
taken these terms disappear.

\end{abstract}

\newpage

\pagestyle{plain}

\setcounter{page}{1}

\newcommand{\ft}[2]{{\textstyle\frac{#1}{#2}}}
\newcommand{\ii}{\mathrm{i}}
\newcommand{\dd}{{\mathrm{d}}}
\newcommand{\nnb}{\nonumber}

\section{Introduction}

Integrability has been the driving force behind the recent years'
progress in the study of the spectral problem of
the $AdS_5/CFT_4$ correspondence~\cite{Minahan:2002ve,Beisert:2003tq,
Bena:2003wd}. Integrability is conjectured to hold in all sectors to all
loop orders~\cite{Beisert:2003tq,Beisert:2005fw}
and impressive tests involving quantities extrapolating
from weak to strong coupling have been
performed~\cite{Beisert:2006ez,
Beisert:2006ib,Basso:2007wd,Bajnok:2008bm}.

Recently a novel explicit example of a gauge/string duality of type
$AdS_4/CFT_3$ has emerged~\cite{Aharony:2008ug} and one could hope
that integrability would play an equally important role there. So
far in the $AdS_4/CFT_3$ correspondence integrability is at a much
less firm setting. The gauge theory dilatation generator has been
proved to be integrable in the scalar sector at leading two-loop
order~\cite{Minahan:2008hf,Bak:2008cp} and the string theory has
been proved to be classically integrable 
in certain subsectors~\cite{Stefanski:2008ik,Arutyunov:2008if,Gomis:2008jt}. 
Investigations probing
integrability at the quantum level of the string theory 
have been carried out in various regimes such as the
BMN limit~\cite{Nishioka:2008gz,Gaiotto:2008cg,Grignani:2008is}, the
giant magnon regime~\cite{Grignani:2008is,Grignani:2008te} and the
near BMN and near flat-space
limits~\cite{Astolfi:2008ji,Kreuzer:2008vd}. There exist conjectures
about integrability of the full $AdS_4/CFT_3$ system  in all sectors to all
loops~\cite{Gromov:2008qe} and a number of tests have come out
affirmative~\cite{Astolfi:2008ji,Ahn:2008aa,McLoughlin:2008ms,McLoughlin:2008he} but certain
problems still seem to require
resolution~\cite{McLoughlin:2008he}.

The spectral information only constitutes one part of the information
encoded in the gauge and string theory. Eventually, one would like to
go beyond the spectral problem and study interacting string theory respectively
non-planar gauge theory. A widespread expectation is that integrability
cannot persist beyond the planar limit. In reference~\cite{Beisert:2003tq}
a way to characterize and quantify the deviation from integrability
was presented for ${\cal N}=4$ SYM. In this case one observed at the
planar level some a priori unexpected
degeneracies in anomalous dimensions
 between certain pairs of operators with opposite
parity. These degeneracies could be explained by the existence of an
extra conserved charge and thus eventually by the integrability of
the theory. When non-planar corrections were taken into account
these degeneracies were found to disappear. Notice, however, that
the degeneracies observed at planar one-loop order persisted when
planar higher loop corrections were taken into account. This
observation was in fact the seed that led to the conjecture about
all loop integrability of ${\cal N}=4$ SYM~\cite{Beisert:2003tq}.

In the present paper we will study non-planar corrections to
${\cal N}=6$ superconformal Chern--Simons--matter theory, the three-dimensional
field theory entering the $AdS_4/CFT_3$ correspondence, in order to
investigate whether one observes a similar
lifting of spectral degeneracies related to integrability
when one goes beyond the planar level. Our investigations will be carried
out in the $SU(2)\times SU(2)$ sector at two-loop level and will thus not
rely on or involve any conjectures.

Using a method based on effective
vertices we will derive the full two-loop
dilatation generator in this sector involving all non-planar corrections.
For short operators the action of this dilatation generator can easily
be written down, resulting in a mixing matrix of low dimension which can
be diagonalized explicitly.\footnote{For ${\cal N}=4$ SYM, explicit diagonalization 
at the non--planar level for a range of operators of this type was carried out in 
\cite{Beisert:2003tq}, see also \cite{Bellucci:2004ru}.} Another type of operators 
for which the mixing matrix can easily be written down are BMN--type
operators~\cite{Berenstein:2002jq} which contain
a large (infinite) number of background fields and a small (finite) number of
excitations. We will look into the nature of the BMN quantum
mechanics~\cite{Beisert:2002ff}
of ${\cal N}=6$ superconformal Chern--Simons--matter theory and will find that in
the BMN scaling limit the two-loop ${\cal N}=6$ theory resembles
the one loop ${\cal N}=4$ SYM theory.
Away from the scaling limit the ${\cal N}=6$
dilatation generator has additional terms.
The mixing problem
of the BMN limit of ${\cal N}=4$ SYM was never solved beyond the planar
limit even
perturbatively in $\frac{1}{N}$ due to
complications arising from huge degeneracies in the planar
spectrum~\cite{Freedman:2003bh}. A third type of operators one could dream of
studying beyond the planar limit are operators dual to
spinning strings.
Such operators typically
contain $M$ excitations and $J$ background fields where
$J,M\rightarrow \infty$ with $\frac{M}{J}$ finite. For such operators, however,
acting
with the dilatation generator involves evaluating infinitely many
terms and writing down the dilatation generator exactly seems intractable.
In reference~\cite{Casteill:2007td} it was suggested that non-planar
corrections to operators dual to
spinning strings could be treated using a coherent state formalism.

Non-planar effects in
${\cal N}=6$ superconformal Chern--Simons--matter theory should reflect
interactions in the dual type IIA string theory. Directly comparable quantities
are, however, not immediate to write down, not least because the $AdS_4/CFT_3$
duality implies the following relation between the string coupling
constant and the gauge theory parameters~\cite{Aharony:2008ug}
\beq
g_s=\frac{\lambda^{5/4}}{N}.
\eeq
This should be compared to the similar relation for ${\cal N}=4$ SYM that
took the form
 $g_s=\frac{\lambda}{N}$ which at least gave the hope that interacting
BMN string states could be studied by perturbative gauge theory computations.
The comparison between the perturbative non-planar gauge theory and
the interacting string theory, described in terms of
light
cone string field theory on a plane wave, however, remained
inconclusive. For a recent review, see~\cite{Grignani:2006en}.
It is thus primarily
 with the purpose of investigating the role of integrability
beyond the planar limit and the structural similarities and differences
between ${\cal N}=4$ SYM and ${\cal N}=6$ superconformal Chern--Simons--matter
theory
that we engage into the present investigations.

We start in section~\ref{summary}
by giving an ultra-short summary of ${\cal N}=6$ superconformal
Chern--Simons--matter theory, i.e.\ ABJM theory.
Subsequently  in section~\ref{derivation1} we derive
the full two-loop
dilatation generator in the $SU(2)\times SU(2)$ sector,
deferring the details to Appendix~\ref{derivation2}. After a
short discussion of the structure of the dilatation generator in
section~\ref{structure} we explain in section~\ref{charges} the relation
between planar degeneracies and conserved charges.
Then
we proceed to apply the dilatation generator to respectively
short operators in section~\ref{short} and BMN operators in
section~\ref{BMNsection}.
Finally,
section~\ref{conclusion} contains our conclusion.

\section{ABJM theory \label{summary}}
Our notation will follow that of references~\cite{Benna:2008zy,Bak:2008cp}.
ABJM theory is a three-dimensional superconformal Chern--Simons--matter
theory with gauge group $U(N)_k\times U(N)_{-k}$ and $R$-symmetry group
$SU(4)$. The parameter $k$ denotes the Chern--Simons level.
The fields of ABJM theory consist of gauge fields $A_m$ and $\bar{A}_m$,
complex scalars $Y^I$ and Majorana spinors $\Psi_I$, $I\in \{1,\ldots 4\}$.
The two gauge fields belong to the adjoint representation of the two
$U(N)$'s. The scalars $Y^I$ and the
spinors $\Psi_I$ transform in the
$N\times \bar{N}$ representation of the gauge group and in the fundamental
and anti-fundamental representation of $SU(4)$ respectively. For our
purposes it proves convenient to write the scalars and spinors explicitly
in terms of their $SU(2)$ component fields, i.e.~\cite{Benna:2008zy}
\bea
Y^I &= &\{Z^A,W^{\dg A}\},
\hspace{0.7cm} Y^\dg_I=\{Z_A^\dg,W_A\},\nonumber \\
\Psi_I&=& \{\epsilon_{AB}\,\xi^B\, e^{i\pi/4},
\epsilon_{AB}\,\omega^{\dg B}\, e^{-i\pi/4},\}, \nonumber\\
\Psi^{I \dg}& = &\{-\epsilon^{AB}\,\xi_B^\dg\, e^{-i\pi/4},
-\epsilon^{AB}\,\omega_B\, e^{i\pi/4}
\}, \nonumber
\eea
 where now $A,B\in \{1,2\}$.
Expressed in terms of these fields the action reads
\bea
S &= &\int d^3x \left [\frac{k}{4\pi} \epsilon^{m n p} \Tr (
A_m \partial_n A_p+\frac{2i}{3} A_m A_n A_p  )-
\frac{k}{4\pi} \epsilon^{m n p} \Tr (
\bar{A}_m \partial_n \bar{A}_p+\frac{2i}{3} \bar{A}_m \bar{A}_n
\bar{A}_p  ) \nonumber
\right. \\
&& \left.
- \Tr ( {\cal D}_m Z)^\dg {\cal D}^m Z-\Tr ({\cal D}_m W)^\dg {\cal D}^m W
+i \Tr \xi^\dg {\cal D}\hspace{-0.3cm}\slash
\hspace{0.13cm}
\xi +i\Tr \omega^\dg {\cal D}\hspace{-0.3cm}\slash
\hspace{0.13cm}\omega  -V^{ferm}-V^{bos}\right]. \nonumber
\eea
Here the covariant derivatives are defined as
\beq
{\cal D}_m Z^A = \partial_m  Z^A + i A_m Z^A-i Z^A \bar{A}_m,
\hspace*{0.7cm}{\cal D}_m W_A = \partial_m  W_A + i \bar{A}_m W_A-i W_A A_m,
\eeq
and similarly for ${\cal D}_m \xi^B$ and ${\cal D}_m \om_B$.
The bosonic as well as the fermionic potential can be separated into
D-terms and F-terms which read
\beq \nonumber
\begin{split} 
V^{ferm}_D =&
\frac{2\pi i}{k} \Tr \left[
(\xi^A \xi_A^\dg\!-\!\om^{\dg A} \om_A)(Z^BZ_B^\dg\!-\!W^{\dg B}W_B)
\!-\!
(\xi_A^\dg \xi^A\!-\!\om_A \om^{\dg A})(Z_B^\dg Z^B\!-\!W_BW^{\dg B} )
\right]  \\
 &
+\frac{4\pi i}{k} \Tr\left[
(\xi^A Z_A^\dg\!-\!W^{\dg A} \om_A)(Z^B \xi_B^\dg\!-\!\om^{\dg B} W_B)
\!-\!
(Z_A^\dg \xi^A\!-\!\om_A W^{\dg A})(\xi_B^\dg Z^B\!-\!W_B \om^{\dg B})
\right], 
\end{split}
\eeq
\beq\nonumber
\begin{split}
V_F^{ferm}&=\frac{2\pi}{k} \epsilon_{AC} \epsilon^{BD}\,
\Tr\Big[ 2\xi^A W_B Z^C \om_D\!+\!2\xi^A \om_B Z^C W_D\!+\!Z^A \om_B Z^C \om_D
\!+\!\xi^A W_B \xi^C W_D\Big] \\
& 
\!\!\!\!+\!\frac{2\pi}{k} \epsilon^{AC}\epsilon _{BD}\,
\Tr \left[ 2 \xi^\dg_A W^{\dg B} Z_C^\dg\om^{\dg D}
\!+\!2\xi_A^\dg \om^{\dg B} Z_C^\dg W^{\dg D}
\!+\!Z_A^\dg \om^{\dg B} Z_C^\dg \om^{\dg D}\!+\!\xi_A^\dg W^{\dg B}\xi_C^\dg
W^{\dg D}\right],
\end{split}
\eeq
\beq
\begin{split}
V_D^{bos}=
\left(\frac{2\pi}{k}\right)^2
\Tr &\left[
\left(Z^A Z_A^\dg + W^{\dg A} W_A\right)
\left(Z^BZ_B^\dg-W^{\dg B}W_B\right)
\left(Z^C Z_C^\dg-W^{\dg C}W_C\right)\right.\\
&+
\left(Z_A^{\dg} Z^A + W_A W^{\dg A}\right)
\left(Z_B^\dg Z^B -W_B W^{\dg B}\right)
\left(Z_C^\dg Z^C - W_C W^{\dg C}\right) \\
& - 2 Z_A^\dg \left(Z^BZ_B^\dg-W^{\dg B}W_B \right) Z^A
\left(Z_C^\dg Z^C - W_C W^{\dg C}\right) \\
&\left. -2 W^{\dg A}\left(Z_B^\dg Z^B -W_B W^{\dg B}\right) W_A
\left(Z^C Z_C^\dg-W^{\dg C}W_C\right) \right]
\end{split}
\eeq
and
\beq
\begin{split}
V_F^{bos}= -\left(\frac{4\pi}{k} \right)^2
\Tr &\left[ W^{\dg A} Z_B^\dg W^{\dg C} W_A Z^B W_C
-W^{\dg A} Z_B^\dg W^{\dg C}W_C Z^B W_A \right. \\
&  \left.
+Z_A^\dg W^{\dg B} Z_C^\dg Z^A W_B Z^C-Z_A^\dg W^{\dg B} Z_C^\dg Z^ C W_B Z^A
\right]. \label{VFbos}
\end{split}
\eeq
Introducing a 't Hooft parameter for the theory
\beq
\lambda=\frac{4\pi N}{k},
\eeq
one can  consider the 't Hooft limit
\beq
N\rightarrow \infty, \hspace{0.7cm} k\rightarrow \infty,
\hspace{0.7cm} \lambda\,\, \mbox{ fixed.}
\eeq
Furthermore, the theory has a double expansion in $\lambda$ and $\frac{1}{N}$.
In this paper we will be interested in studying non-planar effects
for anomalous dimensions at the leading two-loop level.

\section{The derivation of the full dilatation generator \label{derivation1}}

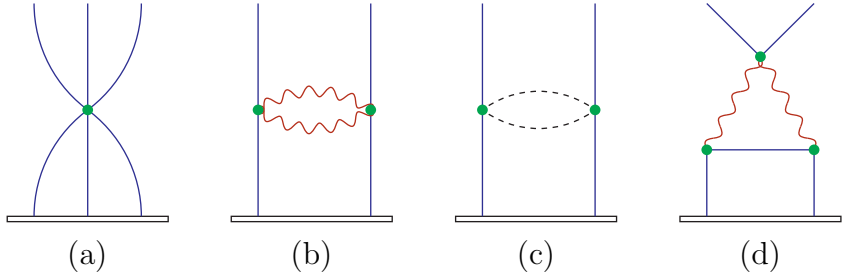
\begin{figure}[t]
\begin{center}
\begin{picture}(80,140)(0,0)
\put(0,10){
\BBox(0,10)(60,8)
\SetColor{Blue}
\Line(30,10)(30,90)
\CArc(60,10)(50,125,180)
\CArc(0,10)(50,0,55)
\CArc(60,90)(50,180,235)
\CArc(0,90)(50,-55,0)
\SetColor{Green}
\Vertex(30,50){2}
\Text(30,-5)[c]{(a)}
}
\end{picture}
\begin{picture}(80,140)(0,0)
\put(0,10){
\BBox(0,10)(60,8)
\SetColor{Blue}
\Line(10,10)(10,90)
\Line(52,10)(52,90)
\SetColor{BrickRed}
\PhotonArc(31,22)(35,51,129){2}{6}
\PhotonArc(31,78)(35,-129,-51){-2}{6}
\SetColor{Green}
\Vertex(10,50){2}
\Vertex(52,50){2}
\Text(30,-5)[c]{(b)}
}
\end{picture}
\begin{picture}(80,140)(0,0)
\put(0,10){
\BBox(0,10)(60,8)
\SetColor{Blue}
\Line(10,10)(10,90)
\Line(52,10)(52,90)
\SetColor{Black}
\DashCArc(31,22)(35,51,129){2}
\DashCArc(31,78)(35,-129,-51){2}
\SetColor{Green}
\Vertex(10,50){2}
\Vertex(52,50){2}
\Text(30,-5)[c]{(c)}
}
\end{picture}
\begin{picture}(80,140)(0,0)
\put(0,10){
\BBox(0,10)(60,8)
\SetColor{Blue}
\Line(10,10)(10,35)
\Line(50,10)(50,35)
\Line(10,35)(50,35)
\Line(30,70)(10,90)
\Line(30,70)(50,90)
\SetColor{BrickRed}
\Photon(10,35)(30,70){2}{4}
\Photon(50,35)(30,70){-2}{4}
\SetColor{Green}
\Vertex(30,70){2}
\Vertex(10,35){2}
\Vertex(50,35){2}
\Text(30,-5)[c]{(d)}
}
\end{picture}
\caption{The four types of two--loop diagrams contributing to anomalous dimensions. For operators
in the $SU(2)\times SU(2)$ sector diagrams in class (d) do not contribute.}\label{Figure}
\end{center}
\end{figure}

In~\cite{Minahan:2008hf,Bak:2008cp} an expression for
the planar dilatation generator
acting on operators of the type
\beq
{\cal O}=\Tr(Y^{A_1} Y_{B_1}^\dg Y^{A_2} Y_{B_2}^\dg\ldots Y^{A_L} Y_{B_L}^\dg),
\eeq
where $A_i,B_i \in \{1,2\}$
was derived and proved to be identical to the Hamiltonian of an integrable
alternating $SU(4)$ spin chain.

Here we will restrict ourselves to considering scalar operators belonging to
a $SU(2)\times SU(2)$ sub-sector i.e.\ operators of the following type
\beq
\label{operators}
{\cal O}= \Tr\left(Z^{A_1} W_{B_1} \ldots Z^{A_L} W_{B_L} \right),
\eeq
and their multi-trace generalizations. For this class of operators
we wish to derive the full dilatation generator including non-planar
contributions. In order to do so we employ
the method of effective vertices from reference~\cite{Beisert:2002bb}.
An effective vertex is a vertex which encodes the combinatorics of a given
type of Feynman diagram. For instance,
the scalar D-terms give rise to the following effective vertex
contributing to the dilatation generator acting on operators of the
type given in eqn.~\rf{operators}
\beq \label{VD}
\begin{split}
\left(V_D^{bos}\right)^{eff}= \gamma \, :\,\Tr &\left[
\left(Z^A Z_A^\dg + W^{\dg A} W_A\right)
\left(Z^BZ_B^\dg\!-\!W^{\dg B}W_B\right)
\left(Z^C Z_C^\dg\!-\!W^{\dg C}W_C\right)\right.\\
& +\left(Z_A^{\dg} Z^A + W_A W^{\dg A}\right)
\left(Z_B^\dg Z^B \!-\!W_B W^{\dg B}\right)
\left(Z_C^\dg Z^C \!-\! W_C W^{\dg C}\right) \\
& \!-\! 2 Z_A^\dg \left(Z^BZ_B^\dg\!-\!W^{\dg B}W_B \right) Z^A
\left(Z_C^\dg Z^C \!-\! W_C W^{\dg C}\right) \\
& \left. \!-\!2 W^{\dg A}\left(Z_B^\dg Z^B \!-\!W_B W^{\dg B}\right) W_A
\left(Z^C Z_C^\dg\!-\!W^{\dg C}W_C\right) \right] :
\end{split}
\eeq
where each daggered field is supposed to be contracted with a field
inside ${\cal O}$, the omissions of self-contractions of the vertex being
encoded in the symbol $:\,\, :$ .
All contractions of $(V_D^{bos})^{eff}$ with the
operator ${\cal O}$ multiply the same Feynman integral whose value we
denote as $\gamma$. The relevant integral is represented by the
Feynman diagram in Fig~1a.
The dilatation generator also gets contributions from the bosonic
$F$-terms, gluon exchange
(Fig.~1b), fermion exchange (Fig.~1c) and scalar self
interactions~\cite{Minahan:2008hf,Bak:2008cp}. Notice, however,
that for operators
belonging to the $SU(2)\times SU(2)$ sector there are no contributions
involving paramagnetic interactions (Fig.~1d).
If things
work as in ${\cal N}=4$ SYM the contribution from
the  D-terms in the sixth order scalar potential should cancel against
contributions from gluon exchange, fermion exchange and self-interactions to
all orders in the genus expansion. We show explicitly in Appendix A that
this is indeed the case. We thus have that the full two-loop dilatation
generator takes the form
\beq \label{normalVFbos}
D=:V_F^{bos}:
\eeq
It is easy to see that the dilatation generator vanishes when acting
on an operator
consisting of only two of the four fields from the $SU(2)\times SU(2)$ sector.
Accordingly we will denote two of the fields, say
$Z_1$ and $W_1$, as
background fields and $Z_2$ and $W_2$ as excitations. It is likewise easy
to see that operators with only one type of excitations, say $W_2$'s, form
a closed set under dilatations. For operators with only
$W_2\:$-excitations the dilatation generator takes the form
\beq
\begin{split}
D= -\left(\frac{4\pi}{k} \right)^2
:\Tr &\left[ W^{\dg 2} Z_1^\dg W^{\dg 1} W_2 Z^1 W_1
-W^{\dg 2} Z_1^\dg W^{\dg 1}W_1 Z^1 W_2 \right. \\
&  \left.
\ +W^{\dg 1} Z_1^\dg W^{\dg 2} W_1 Z^1 W_2
-W^{\dg 1} Z_1^\dg W^{\dg 2}W_2 Z^1 W_1
\right]:
\label{oneexcitation}
\end{split}
\eeq
In the case of two different types of excitations, i.e.\ both $W_2$'s and
$Z_2$'s, the dilatation generator has
16 terms. It appears from the one in~\rf{oneexcitation} by adding  similar terms
with 1 and 2 interchanged and subsequently adding the same operator with
$Z$ and $W$ interchanged. In both cases $D$ is easily seen to reduce to the
one
of~\cite{Minahan:2008hf,Bak:2008cp}
in the
planar limit
\beq
D_{planar}\equiv \lambda^2 D_0= \lambda^2 \sum_{k=1}^{2L} (1-P_{k,k+2}),
\eeq
where $P_{k,k+2}$ denotes the permutation between sites $k$ and $k+2$ and
$2L$ denotes the total number of fields inside an operator.
As explained in~\cite{Minahan:2008hf,Bak:2008cp} this is the Hamiltonian
of two  Heisenberg magnets living respectively on the odd and
the even sites of a spin chain. The two magnets are coupled via the
constraint that the total momentum of their excitations should vanish
which is needed to ensure the cyclicity of the trace.

\section{The structure of the dilatation generator \label{structure}}

As proved in the previous section and in Appendix~\ref{derivation2} 
the two-loop
dilatation generator in the $SU(2)\times
SU(2)$ sector takes the form given in eqn.~\rf{normalVFbos}. When acting on
a given operator we have to perform three contractions as dictated by the
three hermitian conjugate fields. It is easy to see that by acting with
the dilatation generator one can change the number of traces in a given
operator by at most two.\footnote{Acting with the dilatation generator involves
performing three contractions. Performing the first of these does not change
the number of traces. Each of the subsequent contractions on the other hand
can lead to an increase or decrease of the trace number by one.}
More precisely, the two loop dilatation generator has
the expansion
\beq
D= \lambda^2
\left( D_0+\frac{1}{N} D_+ +\frac{1}{N} D_- +\frac{1}{N^2}D_{00}+
\frac{1}{N^2} D_{++}+\frac{1}{N^2} D_{--}\right).
\label{Dexpansion}
\eeq
Here $D_+$ and $D_{++}$ increase the number of traces by one and
two respectively
and $D_-$ and $D_{--}$ decrease the number of traces by one and two.
 Finally, $D_0$ and $D_{00}$ do not change the number of
traces. We notice that in ${\cal N}=4$ SYM the two-loop dilatation
generator in the $SU(2)$ sector
has a similar expansion~\cite{Beisert:2003tq} whereas the
most studied, one-loop dilatation generator involves only  two contractions
and does not contain any $\frac{1}{N^2}$
terms~\cite{Kristjansen:2002bb,Constable:2002hw,Beisert:2002bb,Constable:2002vq}.
Let us assume that
we have found an eigenstate of the planar dilatation generator $D_0$, i.e.
\beq
D_0 |{\cal O}\rangle = E_{\cal O} |{\cal O} \rangle,
\eeq
and let us treat the terms sub-leading in $\frac{1}{N}$ 
as a perturbation. First, let us assume that there are no degeneracies
between $n$-trace states and $(n+1)$-trace states
in the spectrum or that the perturbation has no matrix elements between
such degenerate states.
 If that is the case we can proceed by using non-degenerate
quantum mechanical perturbation theory. Clearly, the leading $\frac{1}{N}$
terms do not have any diagonal components so the energy correction
for the state $|{\cal O}\rangle$ reads:
\beq
\delta E_{\cal O}= \frac{\lambda^2}{N^2}\,\sum_{{\cal K}\neq {\cal O}}
\frac{\langle {\cal O} | D_+ + D_-| {\cal K}\rangle
\langle{\cal K}|D_+ + D_-|{\cal O}\rangle}{E_{\cal O}-E_{\cal K}}
+\frac{\lambda^2}{N^2}\, \langle {\cal O}| D_{00}| {\cal O}\rangle.
\eeq
If there are degeneracies between $n$-trace states and $(n+1)$-trace
states we have to diagonalize the
perturbation in the subset of degenerate states and the corrections
will typically be of order $\frac{1}{N}$.

\section{Planar parity pairs, conserved charges and integrability
\label{charges}}
In the previous sections we derived the two--loop non--planar dilatation generator for the
$SU(2)\times SU(2)$ sector and analyzed its structure. From the work of \cite{Minahan:2008hf,Bak:2008cp}
we know that the planar part of the dilatation generator
can be identified as the Hamiltonian for an integrable
$SU(2)\times SU(2)$ spin chain. It is then interesting to ask what happens to
 integrability once
non-planar corrections are taken into account. One approach to answering this question is to
consider \emph{planar parity pairs}, as we will now review.

As part of their analysis of the dilatation generator of ${\cal N}=4$ SYM, the authors of \cite{Beisert:2003tq}
considered its action on short scalar operators. They observed an a priori unexpected degeneracy
in the resulting spectra, between operators with the same trace structure but opposite \emph{parity}, where
the latter is defined as the operation that reverses the order of all generators within each trace (in
other words, complex conjugation of the gauge group
generators)~\cite{Doikou:1998jh}.
Parity commutes with the action of the dilatation
generator (and is thus a conserved quantity), therefore one expects that the various operators will organize
themselves into distinct sectors according to their (positive or negative) parity.
Positive and negative parity  sectors do not mix with each other and there is no reason to expect 
any relation between their spectra.

However, in \cite{Beisert:2003tq} it was observed that every time there exist operators, which have  the same trace
structure and belong to the same global $SO(6)$ representation but have opposite parity, their \emph{planar}
anomalous dimensions turn out to be equal. This degeneracy could be very simply understood
as a consequence of parity symmetry and
planar integrability: Recall that one of the hallmarks of integrability is the existence
of a tower of commuting conserved charges $Q_n$ (the hamiltonian $Q_2$ being just one of them). For the ${\cal N}=4$ SYM
spin chain there exists such a charge $Q_3$ which (being conserved) commutes with the dilatation generator
but \emph{anticommutes} with the operation of parity. This clearly implies
the existence of pairs of operators with opposite parity and equal anomalous
dimension at the planar level.
Thus  planar integrability manifests itself in the spectrum of short operators through the
appearance of degeneracies between planar parity pairs. Moving beyond planar level, it was observed in
\cite{Beisert:2003tq} that all these degeneracies are lifted: There is no
apparent relation between the
different parity sectors in the spectrum of the non--planar dilatation
generator. This was taken as an
indication (though by no means a proof) that integrability is lost once one considers non--planar
corrections. In this connection, it is worth noticing that the
degeneracies observed at planar one-loop order remain when planar
higher loop
corrections are taken into account~\cite{Beisert:2003tq}.

 Returning to  ${\cal N}=6$ ABJM theory, it is interesting to ask whether the same pattern
of planar degeneracies which are lifted at the non--planar level arises in
the present context. We begin by defining a parity operation which inverts the order of all generators
within each trace, for example:
\beq
\Tr\left[Z_1W_1Z_1W_2Z_2W_1\right]\;
\longrightarrow 
\Tr\left[W_1Z_2W_2Z_1 W_1Z_1\right]=
\Tr\left[Z_1W_1Z_1W_1Z_2W_2\right].
\eeq
Obviously, the Hamiltonian of the $SU(2)\times SU(2)$ spin chain is parity
symmetric. Furthermore,
from the work of~\cite{Minahan:2008hf,Bak:2008cp} we know that the conserved
charges of the $SU(2)\times SU(2)$ spin chain are nothing but the sum of the
charges of the two $SU(2)$ Heisenberg spin chains. In particular,
the third
charge $Q_3$ again anti-commutes with parity while commuting with the
Hamiltonian.

Hence we conclude that we should expect to see parity pairs in the planar part of the spectrum. Furthermore,
the intuition gained from ${\cal N}=4$ SYM points to these degeneracies being broken once non--planar
corrections are taken into account. In the following section, by explicitly considering the action of the dilatation
generator on a series of short operators, we will see that both these expectations are confirmed.

\section{Short Operators \label{short}}

In this section we determine non-planar corrections to 
a number of short operators. This is done by explicitly computing 
and diagonalizing the
mixing matrix (aided by {\tt GPL Maxima} as well as {\tt Mathematica}).

\subsection{Operators with only one type of excitation }

Operators with only one type of excitation can, at the planar level, be
described in terms of just a single Heisenberg spin chain and behave at
the leading two--loop level very similarly
to their ${\cal N}=4$ SYM cousins at one--loop level. Notice, however, that
once one goes
beyond the planar limit the dilatation generator has novel $\frac{1}{N^2}$
terms.
The simplest set of operators for which one observes degenerate parity
pairs as well as non-trivial mixing between operators with different number
of traces
consists of operators of length 14 with three excitations. There are
in total 17 such non-protected operators. Notice that
due to the absence of the trace condition of ${\cal N}=4$ SYM, for which
the gauge group is $SU(N)$, there are more operators here than the naive
generalizations of the ${\cal N}=4$ SYM ones.
Among the
non-protected operators there are only 8 which are not descendants and which we
list below. (To improve readability we suppress the background $Z_1$ fields.)
Notice that only ${\cal O}_1$, ${\cal O}_3$ and ${\cal O}_6$ have analogues in
${\cal N}=4$ SYM.

\beq
\begin{split}
\CO_1&=\Tr([ W_1 W_1, W_1  W_2] W_1 W_2 W_2 )\\
\CO_2&=  \Tr(W_1 ) \Tr( W_1 [W_1, W_2]  W_1 W_2 W_2)\\
\CO_3&=2 \Tr( W_1  W_1  W_1  W_1  W_2  W_2  W_2)
- 3\Tr( W_1 W_2  W_2 W_1 W_1 W_1 W_2) \\
&- 3\Tr ( W_1  W_2 W_1  W_1 W_1  W_2 W_2)
+2 \Tr( W_1 W_2 W_1 W_2 W_1 W_1 W_2)\\
&+ 2 \Tr( W_1 W_1 W_2 W_1 W_1 W_2 W_2)\\
\CO_4&=4 (2\! +\! \sqrt5) \Tr(W_2) \Tr( W_1 W_1 W_1  W_2  W_1 W_2)
\!-\! 2(1\! +\! \sqrt5) \Tr(W_2)
\Tr(W_1 W_1 W_1 W_1 W_2 W_2)\\
&- 2 (3 \!+\! \sqrt5)
\Tr(W_2) \Tr( W_1 W_1 W_2 W_1 W_1 W_2)
\!+\!(3 \!+\! \sqrt5)
\Tr(W_1 ) \Tr(W_1 W_1 W_2 W_1 W_2 W_2)\\
&+(3 \!+\! \sqrt5)\Tr(W_1)
\Tr(W_1 W_2 W_1 W_1  W_2 W_2)
\!-\! 2 \Tr(W_1) \Tr(W_1 W_1 W_1 W_2  W_2 W_2)\\
&\!-\! 2 (2\!+\! \sqrt5)
\Tr(W_1) ( W_2 W_1 W_2 W_1 W_2 W_1)\\
\CO_5&=\!- 4 (2 \!-\! \sqrt5)\Tr(W_2)
\Tr( W_1 W_1 W_1  W_2 W_1 W_2)
\!+\!2 (1 \!-\! \sqrt5)  \Tr(W_2)
\Tr( W_1 W_1 W_1 W_1 W_2 W_2)\\
&\!+\!2 (3\! -\! \sqrt5)
\Tr(W_2) \Tr( W_1 W_1 W_2 W_1 W_1 W_2)
\!-\! (3 \!-\! \sqrt5)  \Tr((W_1)
\Tr(W_1 W_1  W_2  W_1 W_2 W_2)\\
&\!-\! (3\! -\! \sqrt5)
\Tr(W_1 ) \Tr( W_1 W_2 W_1 W_1 W_2 W_2)
+2 \Tr(W_1) \Tr( W_1 W_1 W_1 W_2  W_2 W_2)\\
&+2 (2\!-\! \sqrt5)
\Tr(W_1) \Tr( W_2  W_1 W_2  W_1 W_2 W_1)\\
\CO_6&=\Tr(W_1  W_1 ) \Tr( W_1 [ W_2, W_1]  W_2 W_2)
+\Tr( W_1 W_2)\Tr( W_1 W_1 [ W_1, W_2] W_2) \\
\CO_7&=\Tr(W_1 )\Tr( W_1) \Tr( W_1 [ W_2, W_1]  W_2 W_2)
+\Tr(W_2 ) \Tr( W_1) \Tr( W_1 W_1 [ W_1, W_2] W_2)\\
\CO_8&= \Tr(W_2 ) \Tr( W_1 W_1) \Tr( W_1 [W_2, W_1]  W_2)
+\Tr( W_1) \Tr ( W_1  W_2)\Tr( W_1 [ W_1, W_2] W_2)\\
\end{split}
\eeq

\newpage
The associated planar anomalous dimensions (in units of $\lambda^2$),
trace structure
and parity are
\begin{center}
\begin{tabular}{cccc}
Eigenvector & Eigenvalue & Trace structure & Parity\\ \hline
$\CO_1$ &  $5$ & (14) & $-$\\
$\CO_2$ & $6 $ & (2)(12) & $-$\\
$\CO_3$ & $5$ & (14) & $+$  \\
$\CO_4$ &$5+\sqrt{5}$& (2)(12)  & $+$\\
$\CO_5$ & $ 5-\sqrt{5}$ & (2)(12) & $+$\\
$\CO_{6}$& $4$ & (4)(10) & $+$\\
$\CO_{7}$ &$4$ & (2)(2)(10) & $+$\\
$\CO_{8}$ & $6$ & (2)(4)(8) & $+$\\ 
\end{tabular}
\end{center}
where by parity for multi-trace operators we mean the product of
the parity of its single trace components.
The planar anomalous dimensions of ${\cal O}_1$,
${\cal O}_3$ and ${\cal O}_6$ agree (as they should) with those of
the similar operators in ${\cal N}=4$ SYM,
cf.~\cite{Beisert:2003tq}. We have one pair of degenerate single
trace operators with opposite parity, namely the operators ${\cal
O}_1$ and ${\cal O}_3$.\footnote{We also observe a degeneracy between
the negative parity
double trace state ${\cal O}_2$ and the positive parity
triple trace state $\CO_8$ as well as a degeneracy between the double
trace state $\CO_6$ and the triple trace state $\CO_7$ both of positive
parity.
However, states with different numbers of traces can not be connected
via the conserved charge $Q_3$.}

Expressing the dilatation generator in the
basis above and taking into account all non-planar corrections we
get
\small
\bea
\begin{pmatrix}
5\!+\!\frac{15}{N^2}&\hspace{-.1cm}0&\hspace{-.1cm}0&
\hspace{-.1cm}0&\hspace{-.1cm}0&\hspace{-.1cm}0&\hspace{-.1cm}0 & 0\cr
\hspace{-.1cm}\frac{6}{N^2}&\hspace{-.1cm}
 \!\!6\!+\!\frac{24}{N^2}&\hspace{-.1cm}0&\hspace{-.1cm}\,0&\hspace{-.1cm}0
& 0 & 0 & 0\cr
0 & 0 &
\,\!\,\!5\!+\!\frac{35}{N^2}&\hspace{-.1cm}0&\hspace{-.1cm}0& 
-\frac{8}{N}&\hspace{-.1cm}-\frac{4}{N^2}&\hspace{-.1cm}-\frac{2}{N^2}\cr
0& 0& -\frac{\sqrt{5}}{N}&\hspace{-.1cm}\!\!5\!+\!\sqrt{5}\!+
\!\frac{\left(5\!
\sqrt{5}\!+\!35\right)}{N^2}&\hspace{-.1cm}\frac{3\!\sqrt{5}\!-\!5}{N^2}
&\hspace{-.1cm}\!\!\frac{1}{N^2}
&\hspace{-.1cm}0&\hspace{-.1cm}\!\!\frac{2 }{N}\cr
0& 0& -\frac{\sqrt{5}}{N}&\hspace{-.1cm}-\frac{5\!+\!3\!\sqrt{5}}{N^2}&\hspace{-.1cm}
\!\!{5}\!-\!\sqrt{5}\!-\!\frac{5\!\sqrt{5}\!-\!35}{N^2}&
\hspace{-.1cm}-\frac{1}{N^2}&\hspace{-.1cm}\!
\!\,\,0&\hspace{-.1cm}\!\! -\frac{2}{N}\cr
0& 0 &-\frac{20}{N} &\hspace{-.1cm}\!\!\frac{4\sqrt{5}+20}{N^2}&\hspace{-.1cm}
-\frac{20-4\sqrt{5}}{N^2}&\hspace{-.1cm}\!\!4\!+\!\frac{28}{N^2}&
\hspace{-.1cm}\,0&\hspace{-.1cm}0\cr
0& 0& 
-\frac{10}{N^2}&\hspace{-.1cm}\frac{4\sqrt{5}+20}{N}&
\hspace{-.1cm}\frac{4\sqrt{5}-20}{N}&\hspace{-.1cm}0&\hspace{-.1cm}
 \!\!4\!+\!\frac{32}{N^2}&\hspace{-.1cm}-\frac{2}{N^2}\cr
0& 0& 
-\frac{10}{N^2}&\hspace{-.1cm}\!
\!\frac{24\sqrt{5}+40}{N}&\hspace{-.1cm}
\frac{24\sqrt{5}-40}{N}&
\hspace{-.1cm}\frac{8}{N}&\hspace{-.1cm}-\frac{8}{N^2}
&\!6\!+\!\frac{40}{N^2}\,\cr
\end{pmatrix}\;.
\eea
\normalsize
Notice the decoupling of positive and negative parity states and
the presence of numerous $\frac{1}{N^2}$-terms which do not have
analogues in one-loop ${\cal N}=4$ SYM. One observes that 
the states ${\cal O}_1$ and ${\cal O}_2$ are exact eigenstates
of the full dilatation generator with non-planar corrections
equal to
\beq
\delta E_1 =  \frac{15}{N^2}, \hspace{0.7cm}
\delta E_2 =  \frac{24}{N^2}. \hspace{0.7cm}
\eeq
For the remaining operators we observe that all matrix elements between 
degenerate states vanish.
Thus the leading non-planar corrections to the anomalous dimensions
can be found using second order non-degenerate perturbation theory.
The results read
\bea \delta E_3 &= & \frac{195}{N^2}, \hspace{0.7cm}
\delta E_4 = \frac{115+37\sqrt{5}}{N^2}, \hspace{0.7cm} \nonumber\\
\delta E_5 &=& \frac{115-37\sqrt{5}}{N^2}, \hspace{0.7cm}
\delta E_6  =  -\frac{132}{N^2}, \\
\delta E_7 & =& \frac{32}{N^2}, \hspace{0.7cm} \delta E_8
= -\frac{120}{N^2}.\nonumber \eea
We observe that all degeneracies found at the planar level get
lifted when non-planar corrections are taken into account. This in
particular holds for the degeneracies between the members of the
planar parity pair $({\cal O}_1,{\cal O}_3)$. Notice that whereas
the planar eigenvalues of the operators ${\cal O}_1$, ${\cal O}_3$
and ${\cal O}_6$ are identical to those of their ${\cal N}=4$ SYM
cousins the non-planar corrections are not.


\subsection{Operators with two types of excitations}

An operator with two excitations of different type corresponds in
spin chain language to the situation where each of the two coupled
spin chains has one excitation. Such an operator does not
immediately have an analogue in ${\cal N}=4$ SYM. (One can indeed
consider scalar ${\cal N}=4$ SYM operators with two types of
excitations $\Phi$ and $\Psi$ on a background of $Z$ fields but
these operators should be organized into representations of $SO(6)$,
and not of $SU(2)\times SU(2)$ as here, and thus always come in
symmetrized or antisymmetrized versions.)

\subsubsection{Length 8 with 2 excitations \label{twoexcitations}}
Let us analyze the simplest multiplet of operators with two excitations
of different types that exhibit some
of the above
mentioned
non-trivial features of the $\frac{1}{N}$-expansion, operators of length
eight with
one excitation of each type. There are in total 7 such non-protected
operators. The planar non-protected
eigenstates of the two-loop dilatation
generator read
\beq
\begin{split}
\CO_{ 1 }=&\Tr(Z_1 W_1 \{ Z_1 W_2, Z_2 W_1\} Z_1 W_1)-\Tr( W_1 Z_1 \{W_1 Z_2, W_2 Z_1\} W_1 Z_1 )\\
\CO_{ 2 }=& -\Tr(Z_1 W_1 [Z_1 W_2, Z_2 W_1] Z_1 W_1)+\Tr( W_1 Z_1 [W_1 Z_2, W_2 Z_1] W_1 Z_1)\\ 
\CO_{ 3 }=& \Tr(Z_1 W_1 Z_1 W_1)\left[\Tr( Z_1 W_2 Z_2 W_1) - \Tr( W_1 Z_2 W_2 Z_1) \right]\\ 
\CO_{ 4 }=& \Tr(W_1 Z_1) \left[\Tr(W_1 Z_1 W_2 Z_2 W_1 Z_1)- \Tr( Z_1 W_1 Z_2 W_2 Z_1 W_1)\right]\\  
\CO_{ 5 }=& \Tr(W_1 Z_1) \Tr(W_1 Z_1) \left[\Tr(W_2 Z_1 W_1 Z_2) -\Tr(Z_2 W_1 Z_1 W_2 )\right]\\ 
\CO_{ 6 }=& - \Tr(Z_1 W_1 [Z_1 W_2, Z_2 W_1] Z_1 W_1)- \Tr( W_1 Z_1 [W_1 Z_2, W_2 Z_1] W_1 Z_1)\\ 
\CO_{ 7 }=& -\Tr(W_1 Z_1)\left[ \Tr( W_1 Z_1 [W_1 Z_2, W_2 Z_1])+ \Tr(Z_1 W_1 [Z_1 W_2, Z_2 W_1])\right]\\
\end{split}
\eeq
and the associated planar anomalous dimensions (in units of
$\lambda^2$), trace structure and parity are
\begin{center}
\begin{tabular}{cccc}
Eigenvector & Eigenvalue & Trace Structure &Parity\\ \hline
 $\CO_1$ & 8  & (8)&$-$\\
$\CO_2$ & 4 & (8)&$-$\\
 $\CO_3$ &  8 & (4)(4)&$-$\\
$\CO_4$ & 6 & (2)(6)&$-$\\
$\CO_5$ & 8 & (2)(2)(4)&$-$  \\
$\CO_6$ & 4& (8)&$+$\\
$\CO_7$ & 6 & (2)(6)&$+$\\
\end{tabular}
\end{center}
Notice that we have two pairs of degenerate operators with
opposite parity, namely the single trace operators $\CO_2$ and $\CO_6$ and
the double trace operators $\CO_4$ and $\CO_7$.\footnote{The double trace
operators ${\cal O}_4$ and ${\cal O}_7$
can be related via $Q_3$ when letting $Q_3$ act only on the
longer of the two constituent traces of the operators.}

Expressing the dilatation generator in the basis given above and taking into
account all non--planar corrections we get
\beq
\begin{pmatrix}
8&\frac{8}{\:N^2}&\frac{16}{N}&\frac{4}{N}&-\frac{8}{\:N^2}&0&0\cr 
\frac{8}{\:N^2}&\!4\!-\!\frac{12}{N^2}&0&-\frac{2}{N}&-\frac{4}{\:N^2}&0&0 \cr 
\frac{16}{N}&-\frac{8}{N}&8 &0&0&0&0\cr 
0&-\frac{16}{N}&-\frac{8}{\:N^2}&\!6\!-\!\frac{8}{\:N^2}&-\frac{12}{N}&0&0 \cr 
0&\frac{8}{\:N^2}&0&-\frac{12}{N}&\!8\!-\!\frac{8}{\:N^2}&0&0\cr 
0&0&0&0&0&\!4\!+\!\frac{4}{N^2}&\frac{2}{N}\cr 
0&0&0&0&0 &\frac{8}{N}&\!6\!+\!\frac{8}{\:N^2}\cr 
\end{pmatrix}\;.
\eeq
The non-planar corrections for ${\cal O}_6$ and ${\cal O}_7$ can be
found exactly and read
\beq
\delta E_{6,7}= 
\frac{6}{\:N^2}\mp \left(\sqrt{1+\frac{20}{\:N^2}+\frac{4}{\:N^4}}
- 1\right).
\eeq
The corrections to the eigenvalues of the remaining operators we instead
find using perturbation theory as described in section~\ref{structure}.
First we notice that most matrix elements between degenerate states vanish.
The only exception are the matrix elements between the
states $\CO_1$ and $\CO_3$. To find the non--planar correction to the energy
of these states we diagonalize the Hamiltonian in the corresponding subspace
and find
\beq \delta E_{1,3}=\mp  \, \frac{16}{N}. \eeq
For the remaining operators the leading non-planar corrections to
the energy can be found using second order non-degenerate
perturbation theory. The results read
\beq \delta E_2 =  -\frac{28}{N^2}, \hspace{0.7cm}
\delta E_4 =  -\frac{64}{N^2}, \hspace{0.7cm}
\delta E_5 = \frac{64}{N^2}\;. 
\eeq
We again notice that all
degeneracies observed at the planar level get lifted when non-planar
corrections are taken into account. This in particular holds for the
degeneracies between the members of the two parity pairs.

\subsubsection{Length 8 with 3 excitations \label{threeexcitations}}

We now consider operators with three excitations, one of type $Z_2$ and
two of type $W_2$. 
Among this type of operators one finds 7 which are descendants of the 
7 operators considered in the previous section. Of highest weight states
one has the following four planar eigenstates: 
\beq
\begin{split} 
\CO_1=&\Tr( Z_1W_2) \left[\Tr(Z_1 W_1 Z_2 W_2 Z_1 W_1)-\Tr(W_1 Z_1 W_2 Z_2 W_1 Z_1)\right]\\
 &- \Tr(Z_1W_1 ) \left[\Tr(Z_1 W_1 Z_2 W_2 Z_1 W_2)-\Tr(Z_1 W_2 Z_2 W_1 Z_1 W_2)\right]\\
\CO_2=&\Tr(Z_1 W_1 [Z_2 W_1, Z_1 W_2] Z_1 W_2)+\Tr( Z_1 W_2 [Z_1 W_2, Z_2 W_1] Z_1 W_1) \\ 
&+\Tr(Z_1 W_1 [Z_1 W_1, Z_2 W_2] Z_1 W_2)+\Tr( Z_1 W_2 [Z_2 W_2 ,Z_1 W_1] Z_1 W_1)\\ 
\CO_3=&-\Tr(W_2 Z_1 [W_1 Z_1, W_1 Z_2] W_2 Z_1 )+
\Tr( W_1 Z_1 [W_2 Z_2, W_2 Z_1] W_1 Z_1)\\
\CO_4=&\Tr(Z_1W_2 )\left[ \Tr( W_1 Z_1 [W_1 Z_2, W_2 Z_1])+ \Tr(Z_1 W_1 [Z_1 W_2, Z_2 W_1] )\right]\\
&+\Tr(Z_1W_1 ) \left[\Tr( Z_1 W_2 [Z_1 W_1, Z_2 W_2])+\Tr( W_2 Z_1 [W_2 Z_2, W_1 Z_1])\right]
\end{split}
\eeq
Their planar anomalous dimensions (in units of $\lambda^2$), trace structure
and parity are tabulated below.
\begin{center}
\begin{tabular}{cccc}
Eigenvector & Eigenvalue & Trace Structure & Parity\\ \hline
$\CO_1$ & $6 $ &$(2)(6)$  & $-$\\
$\CO_2$ & $6 $ & $(8)$ & $+$\\
$\CO_3$ & $6 $ & $(8)$ & $+$\\
$\CO_4$ & $6 $ & $(2)(6)$ &  $+$\\
\end{tabular}
\end{center}
We observe one planar parity pair with trace structure $(2)(6)$.
The full mixing matrix for this set of states takes the following form:
\beq
\left(\begin{array}{cccc}
\!6\!-\frac{16}{\:N^2}&0&0&0 \cr 
0&\!6\!+\!\frac{12}{\:N^2}&0&0 \cr 
0&0&6-\frac{4}{\:N^2}&-\frac{12}{N} \cr 
0&0&-\frac{4}{N}& 6 \cr 
\end{array}\right)
\eeq
and the exact non-planar corrections to the energy are
\bea
\delta E_1 &=& - \frac{16}{N^2},\hspace{0.7cm} \delta E_2= \frac{12}{N^2}, 
\nonumber \\
\delta E_{3,4}&=& -\frac{2}{N^2}\pm 2\sqrt{\frac{12}{N^2}+\frac{1}{N^4}}.
\eea
Also in this case it turns out that all planar degeneracies are lifted.
Obviously, there is another three-excitation sector with one $W_2$-excitation
and two $Z_2$-excitations. The results for that sector can of course
easily be read off from those of the present one.

\subsubsection{Length 8 with 4 excitations}

Let us turn to the case of operators of length eight with two excitations
of type $W_2$ and two excitations of type $Z_2$. In this sector we find seven
operators which descend from the operators treated in 
section~\ref{twoexcitations} as well as eight operators which descend from
operators with three excitations. The remaining non-protected operators
are
\beq
\begin{split}
\CO_{ 1 }=& - \Tr(Z_1 W_1 Z_1 W_1 Z_2 W_2 Z_2 W_2)+\Tr(W_1 Z_1 W_1 Z_1 W_2 Z_2 W_2 Z_2) \\
&+\Tr(W_2 Z_1 W_2 Z_1 W_1 Z_2 W_1 Z_2)- \Tr(W_1 Z_2 W_1 Z_1 W_2 Z_1 W_2 Z_2)\\ 
\CO_{ 2 }=& \Tr(W_1 Z_2)\left[ \Tr(Z_1 W_2 Z_1 W_1 Z_2 W_2) - \Tr(W_1 Z_1 W_2 Z_1 W_2 Z_2)\right] \\
&+ \Tr(Z_2 W_2)\left[ \Tr(Z_1 W_1 Z_1 W_2 Z_2 W_1)- \Tr(W_2 Z_1 W_1 Z_1 W_1 Z_2)\right]  \\
&+\Tr(Z_1 W_2)\left[ \Tr(Z_1 W_1 Z_2 W_1 Z_2 W_2)-\Tr(W_1 Z_2 W_1 Z_1 W_2 Z_2)\right]\\
& +\Tr(W_1 Z_1)\left[ \Tr(W_1 Z_1 W_2 Z_2 W_2 Z_2) - \Tr(W_2 Z_1 W_1 Z_2 W_2 Z_2)\right]\\ 
\CO_{ 3 }=& \Tr(W_1 Z_1)  \Tr(Z_2 W_2)\left[\Tr(W_1 Z_1 W_2 Z_2) - \Tr(W_2 Z_1 W_1 Z_2)\right]\\ 
&+\Tr(W_1 Z_2) \Tr(Z_1 W_2) \left[\Tr(Z_1 W_1 Z_2 W_2)-\Tr(W_1 Z_1 W_2 Z_2)\right] \\
\CO_{ 4}=& \Tr(Z_1 W_1 \{Z_1 W_1, Z_2 W_2\} Z_2 W_2)+\Tr(Z_2 W_1\{ Z_2 W_1, Z_1 W_2\} Z_1 W_2 )\\
&+\Tr(W_1 Z_1 \{W_1 Z_1, W_2 Z_2\} W_2 Z_2) +\Tr(W_2 Z_1\{ W_2 Z_1, W_1 Z_2\} W_1 Z_2)\\
&- 2 \Tr( W_1 Z_1 \{W_1 Z_2, W_2 Z_1\} W_2 Z_2)- 2 \Tr(Z_1 W_1\{Z_1 W_2, Z_2 W_1\} Z_2 W_2 )\\ 
\CO_{ 5}=&- \Tr(Z_2 W_2)\left[ \Tr([W_2 Z_1, W_1 Z_1] W_1 Z_2)+ \Tr([Z_1 W_1, Z_1 W_2] Z_2 W_1)\right]\\ 
&-\Tr(W_1 Z_2)\left[\Tr([Z_1 W_2, Z_1 W_1] Z_2 W_2) + \Tr([W_1 Z_1, W_2 Z_1] W_2 Z_2)\right]\\
&-\Tr(Z_1 W_2)\left[ \Tr([Z_1 W_1, Z_2 W_1] Z_2 W_2) + \Tr([W_1 Z_2, W_1 Z_1] W_2 Z_2)\right]\\ 
&- \Tr( Z_1 W_1)\left[\Tr( [Z_1 W_2, Z_2 W_2] Z_2 W_1) + \Tr([W_2 Z_1, W_1 Z_2] W_2 Z_2)\right]\\
\CO_{ 6}=& 2\Tr(W_1 Z_1 W_2 Z_2) \Tr(Z_1 W_1 Z_2 W_2) - 
\Tr(W_2 Z_1 W_1 Z_2) \Tr(Z_1 W_1 Z_2 W_2) \\
&- \Tr(W_1 Z_1 W_2 Z_2) \Tr(W_1 Z_1 W_2 Z_2)
\end{split}
\eeq
with planar eigenvalues (in units of $\lambda^2$), trace structure and
parity given
by
\begin{center}
\begin{tabular}{cccc}
Eigenvector & Eigenvalue & Trace Structure & Parity\\ \hline
$\CO_1$ & $4 $ &$ (8)$  & $-$\\
$\CO_2$ & $6 $ & $(2)(6)$ & $-$\\
$\CO_3$ & $8 $ & $(2)(2)(4)$ & $-$\\
$\CO_4$ & $12 $ &$ (8)$  & $+$\\
$\CO_5$ & $6 $ & $(2)(6)$ & $+$\\
$\CO_6$ & $16 $ & $(4)(4)$ & $+$\\
\end{tabular}
\end{center}
We notice one planar parity pair with trace structure $(2)(6)$.
In the subspace of
negative parity operators the dilatation generator reads
\beq
\left(\begin{array}{ccc}
\!4\!-\!\frac{12}{\:N^2}&\frac{12}{N}& \frac{12}{\:N^2} \cr 
\frac{12}{N}&\!6\!& \frac{6}{N} \cr 
\frac{8}{\:N^2}&\frac{24}{N}&\!8\!-\!\frac{\!8\!}{\:N^2} \cr 
\end{array}\right)\;.
\eeq
The leading corrections to the eigenvalues can be found to be
\beq
\delta E_1= -\frac{84}{\:N^2},
\hspace{0.7cm} \delta E_2= -\frac{1728}{\:N^4} ,
\hspace{0.7cm} \delta E_3= \frac{64}{\:N^2}.
\eeq
The mixing matrix in the subspace of positive parity eigenvalues
looks as follows:
\beq
\left(\begin{array}{ccc}
\!12\!-\!\frac{12}{\:N^2}&-\frac{12}{N}& -\frac{8}{N} \cr 
0&\!6\!& -\frac{8}{\:N^2} \cr 
-\frac{72}{N}&0 &\!16\! \cr 
\end{array}\right)\;.
\eeq
For these states we find the following leading corrections:
\beq
\delta E_4= -\frac{156}{\:N^2},\hspace{0.7cm}
\delta E_5 = -\frac{576}{5\:N^4},\hspace{0.7cm}
\delta E_6 = \frac{144}{\:N^2}.
\eeq
Again we see that all planar degeneracies are lifted.\footnote{However, it is worth noting that the resolution
of the degeneracy between ${\cal O}_2$ and ${\cal O}_5$ happens at order $1/N^4$ and would thus not be 
visible purely within second order perturbation theory.}

Summarizing, in all sectors considered we have observed a degeneracy
between operators with similar trace structure but opposite parity -- a
degeneracy which, as explained earlier, could be attributed to the
existence of an extra conserved charge and thus to the integrability
of the planar dilatation generator.
The lift of degeneracies
 can be taken as an indication (but not a proof)
that integrability breaks down beyond the planar level. In any case
the concept of integrability when formulated in terms of spin chains
and their associated conserved charges has to be
reformulated when multi-trace operators are taken into account but
it is clear that some symmetries are lost when we go beyond the
planar limit.


\section{BMN operators \label{BMNsection}}

In the previous section we analyzed the case of short operators in
ABJM theory. Another important class of operators that played a
crucial role in the context of the $AdS_5/CFT_4$ correspondence is that of the
so-called BMN operators \cite{Berenstein:2002jq}. 
It is not difficult to see that BMN operators of ABJM theory can be
constructed analogously to BMN operators of ${\cal N}=4$ SYM
\cite{Berenstein:2002jq}.

In this section we compute non-planar corrections to the anomalous dimensions 
of BMN-type operators in the $SU(2)\times SU(2)$ sector of ABJM
theory~\cite{Nishioka:2008gz, Gaiotto:2008cg,Grignani:2008is}.
We will restrict ourselves to considering BMN operators with two
excitations. There are two types of such operators:\footnote{As
pointed out in \cite{Minahan:2008hf}, these operators resemble
scalar operators in the orbifolds of ${\cal N} = 4$ SYM theory in
four dimensions. Non-planar corrections for operators in the
orbifolded ${\cal N} = 4$ SYM theory have been computed
in~\cite{Bertolini:2002nr, DeRisi:2004bc}.}
\begin{equation}
\CA_l^{{J}_0,J_1,\ldots,J_k}= \tr\!\!\left[Z_2\left(W_1Z_1\right)^l
W_2\left(Z_1W_1\right)^{{J}_0-l}\right]\tr\!\!\left[\left(Z_1W_1\right)^{J_1}\right]
\ldots \tr\!\!\left[\left(Z_1W_1\right)^{J_k}\right], \label{OAB}
\end{equation}
\begin{equation}
\CB_l^{{J}_0,J_1,\ldots,J_k}= \tr\!\!\left[\left(Z_1W_1\right)^lZ_1
W_2\left(Z_1W_1\right)^{{J}_0-l}Z_1W_2\right]\tr\!\!\left[\left(Z_1W_1\right)^{J_1}\right]
\ldots \tr\!\!\left[\left(Z_1W_1\right)^{J_k}\right]. \label{OBB}
\end{equation}
There are in total $J_0+1$ independent operators of type $\CA$ and
$[J_0/2]+1$ independent operators of type $\CB$. The associated bare
conformal dimensions are
\beq \Delta_{\CA}= J_0+\ldots +J_k+1, \hspace{0.7cm}\Delta_{\CB}=
J_0+\ldots+J_k+2. \eeq
In the spin chain language the $\CB$-operators have two excitations
on the same spin chain whereas the $\CA$-operators have one
excitation on each spin chain. As already mentioned, the
$\CA$-operators do not have an analogue in the scalar sector of
${\cal N}=4$ SYM\footnote{This was first pointed out in
\cite{Astolfi:2008ji} from the analysis of the dual string theory
state.} where operators have to organize into representations of
$SO(6)$ (and not into representations of $SU(2)\times SU(2)$ as
here). In ${\cal N}=4$ SYM two--excitation operators always appear in a
symmetrized or anti-symmetrized version.

We wish to study the non-planar corrections to both types of
operators. As in ${\cal N}=4$ SYM we find the set of two--excitation
operators above are closed under the action of the dilatation
generator, i.e. two--excitation operators with the two excitations in
two different traces are never generated when the dilatation
generator acts. In the next two sub-sections we consider separately
the two sets of operators $\CA_l^{{J}_0,J_1,\ldots,J_k}$ and
$\CB_l^{{J}_0,J_1,\ldots,J_k}$.

Introducing $J=J_0+J_1+\ldots+J_k$ we define the BMN limit as the
double scaling limit~\cite{Kristjansen:2002bb,Constable:2002hw}
\beq J\rightarrow \infty, \hspace{0.7cm} N\rightarrow \infty,
\hspace{0.7cm}\lambda' \equiv
\frac{\lambda^2}{J^2},\hspace{0.5cm} g_2=\frac{J^2}{N},
\hspace{0.3cm} \mbox{fixed}. \label{BMN} \eeq
The BMN limit of the ${\cal N}=6$ superconformal Chern--Simons--matter 
theory is expected to correspond to the Penrose limit of the
type IIA string theory on $AdS_4\times CP^3$. The string theory
states dual to the BMN operators $\CA_l^{{J}_0,J_1,\ldots,J_k}$ and
$\CB_l^{{J}_0,J_1,\ldots,J_k}$ have been studied
in~\cite{Grignani:2008is, Astolfi:2008ji}. Notice, however, that due
to different dispersion relations of excitations in the spin chain
and string theory language~\cite{Grignani:2008is} the correct
definition of $\lambda'$ at leading order in a strong coupling
expansion is
$\lambda'=\lambda/J^2$~\cite{Gaiotto:2008cg,Grignani:2008is}.

\subsection{BMN operators with only one type of excitation}
For operators with only one type of excitation the dilatation generator is
given by the expression in eqn.~\rf{oneexcitation}. Using the notation of
eqn.~\rf{Dexpansion} we find
\begin{equation}
D_0\circ\CB_{p}^{J_0,J_1,\ldots,J_k}= -2
\left(\delta_{p\neq J_0}\CB_{p+1}^{J_0,J_1,\ldots,J_k}+\delta_{p\neq
0}\CB_{p-1}^{J_0,J_1,\ldots,J_k} -(\delta_{p\neq 0}+\delta_{p\neq
J_0})\CB_{p}^{J_0,J_1,\ldots,J_k}\right),\label{H0B}
\end{equation}
\beq
\begin{split}
D_+\circ\CB_{p}^{J_0,J_1,\ldots,J_k}= - 4&
\left[\sum_{J_{k+1}=1}^{p-1}\left(\CB_{p-J_{k+1}-1}^{J_0-J_{k+1},J_1,\ldots,J_k,J_{k+1}}
-\CB_{p-J_{k+1}}^{J_0-J_{k+1},J_1,\ldots,J_k,J_{k+1}}\right)\right.\cr
&\left.-
\sum_{J_{k+1}=1}^{J_0-p-1}\left(\CB_{p}^{J_0-J_{k+1},J_1,\ldots,J_k,J_{k+1}}
-\CB_{p+1}^{J_0-J_{k+1},J_1,\ldots,J_k,J_{k+1}}\right) \right]
\label{H+B}
\end{split}
\eeq
and
\beq
\begin{split}
D_-\circ\CB_{p}^{J_0,J_1,\ldots,J_k}= - 4
&\left[\sum_{i=1}^{k}J_i\left(\CB_{J_i+p-1}^{J_0+J_i,J_1,\ldots,
\makebox[0pt]{\,\,\,\,$\times$}J_i, \ldots,J_k}
-\CB_{J_i+p}^{J_0+J_i,J_1,\ldots,
\makebox[0pt]{\,\,\,\,$\times$}J_i, \ldots,J_k}\right.\right.\cr
&\qquad\quad\; \left.-\CB_{p}^{J_0+J_i,J_1,\ldots,
\makebox[0pt]{\,\,\,\,$\times$}J_i, \ldots,J_k}
+\CB_{p+1}^{J_0+J_i,J_1,\ldots, \makebox[0pt]{\,\,\,\,$\times$}J_i,
\ldots,J_k}\right)\bigg]. \label{H-B}
\end{split}
\eeq
The terms resulting from the action of $D_{++}$, $D_{--}$ and $D_{00}$
are rather involved and we have deferred them to Appendix B.

We notice that the form of $D_0$, $D_+$ and $D_-$ are exactly as for
${\cal N}=4$ SYM at one loop order, written down in the same
notation in~\cite{Beisert:2003tq}, except for
the fact that $D_+$ and $D_-$ in the present case have an additional
factor of 2 compared to $D_0$. Thus for this type of operators the
analysis up to order $\frac{1}{N}$ can be directly carried over
from~\cite{Beisert:2003tq}. At order $\frac{1}{N^2}$ one has to take
into account the novel terms $D_{00}$, $D_{++}$ and $D_{--}$
appearing in Appendix~\ref{Boperators}. However, as explained there
once one imposes the
BMN limit defined in eqn.~\rf{BMN} these terms become sub-dominant.
The BMN quantum mechanics is therefore (up to trivial factors of two)
identical to that of ${\cal N}=4$ SYM at one loop level. In particular one
encounters the same problem that the huge degeneracies make the
perturbative treatment of the non-planar corrections intractable.

\subsection{BMN operators with two different types of excitations}
For operators with two different types of excitations the dilatation
generator is given by the expression~\rf{oneexcitation} where we add the
similar terms with 1 replaced by 2 and subsequently add the same operator
with $Z$ and $W$ interchanged. Thus, in this case the dilatation generator
consists of 16 terms. Using the notation of eqn.~\rf{Dexpansion} we find
\beq
D_0\circ\CA_{p}^{J_0,J_1,\ldots,J_k}= -2
\left(\delta_{p\neq J_0}\CA_{p+1}^{J_0,J_1,\ldots,J_k}+\delta_{p\neq
0}\CA_{p-1}^{J_0,J_1,\ldots,J_k} -(\delta_{p\neq J_0}+\delta_{p\neq
0})\CA_{p}^{J_0,J_1,\ldots,J_k}\right),
\eeq
\beq \label{H+}
\begin{split}
D_+\circ\CA_{p}^{J_0,J_1,\ldots,J_k}= -
&\left[4\sum_{J_{k+1}=1}^{p-1}\left(\CA_{p-J_{k+1}-1}^{J_0-J_{k+1},J_1,\ldots,J_k,J_{k+1}}
-\CA_{p-J_{k+1}}^{J_0-J_{k+1},J_1,\ldots,J_k,J_{k+1}}\right)\right. \\
&\left.-4\sum_{J_{k+1}=1}^{J_0-p-1}\left(\CA_{p}^{J_0-J_{k+1},J_1,\ldots,J_k,J_{k+1}}
-\CA_{p+1}^{J_0-J_{k+1},J_1,\ldots,J_k,J_{k+1}}\right)\right.\\
&+2\delta_{p\neq 0}\left(\CA_{0}^{p,J_1,\ldots,J_k,J_0-p}
-\CA_{p}^{p,J_1,\ldots,J_k,J_0-p}\right)\\
&+ 2\delta_{p\neq J_0}\left(\CA_{J_0-p}^{J_0-p,J_1,\ldots,J_k,p}
-\CA_{0}^{J_0-p,J_1,\ldots,J_k,p}\right) \bigg]
\end{split}
\eeq
and
\beq\label{H-}
\begin{split}
D_-\circ\CA_{p}^{J_0,J_1,\ldots,J_k}= - 4
\sum_{i=1}^{k}J_i&\left[(\CA_{J_i+p-1}^{J_0+J_i,J_1,\ldots,
\makebox[0pt]{\,\,\,\,$\times$}J_i, \ldots,J_k}
-\CA_{J_i+p}^{J_0+J_i,J_1,\ldots,
\makebox[0pt]{\,\,\,\,$\times$}J_i, \ldots,J_k})\right.\cr
&-
\left.(\CA_{p}^{J_0+J_i,J_1,\ldots,
\makebox[0pt]{\,\,\,\,$\times$}J_i, \ldots,J_k}
-\CA_{p+1}^{J_0+J_i,J_1,\ldots, \makebox[0pt]{\,\,\,\,$\times$}J_i,
\ldots,J_k})\right]. 
\end{split}
\eeq
The contributions arising from the action of $D_{++}$, $D_{--}$ and $D_{00}$
can be found in Appendix B.
Formally $D_0$, $D_+$ and $D_-$ are similar to the ones one obtains
when applying the one-loop dilatation generator of ${\cal N}=4$ SYM to an
operator containing two different excitations (i.e. $\Psi$ and $\Phi$ in a background
of $Z$'s). The only differences are that the
quantities $D_+$ and $D_-$ in the present case have an additional
factor of 2 compared to $D_0$ and that there appear two Kronecker
$\delta$'s
in $D_+$. However, as already mentioned, in ${\cal N}=4$ SYM operators with two
excitations of different types have to organize into representations of
$SO(6)$
and therefore always come in a symmetrized or anti-symmetrized form.
For symmetrized operators, the last line of eqn.~\rf{H-} vanishes. Taking
the BMN limit we observe as before that the terms $D_{++}$, $D_{--}$
and $D_{00}$ become sub-dominant, cf.\ Appendix~\ref{Aoperators}.

\section{Conclusion \label{conclusion}}

We have derived and studied the full two-loop dilatation generator
in the $SU(2)\times SU(2)$ sector of ${\cal N}=6$ superconformal
Chern--Simons--matter theory. As opposed to what was the case at
leading order in ${\cal N}=4$ SYM theory, the leading order
dilatation generator of ABJM theory implies a mixing not only
between $n$ and $(n+1)$ trace states but also between $n$ and
$(n+2)$ trace states. The latter mixing
 becomes sub-dominant when the BMN limit is considered.

By acting with the dilatation generator on short operators we observed
at the planar level pairs of degenerate
operators belonging to the same representation but having opposite
parity. As in planar ${\cal N}=4$ SYM these degenerate parity pairs could
be explained by the existence of an extra conserved charge, the first
of the tower of conserved charges of the alternating $SU(2)\times SU(2)$
spin chain. When non-planar corrections were taken into account these
degeneracies disappeared indicating (but not
proving) the breakdown of integrability. It would of course be interesting
to investigate the mixing problem for higher representations
of $SU(2)\times SU(2)$ than the
ones considered here to see if other types of symmetries will reveal
themselves. It is clear, however, that
once one allows for mixing between
operators with different number of traces  one
needs to re-think the entire concept of integrability. The simple spin
chain picture breaks down and the concept of local charges becomes
inadequate. In fact, it would be interesting to try to construct
a toy example of what one
could call an integrable model involving splitting and joining
of traces, perhaps along the lines of the simple solvable toy model 
of reference~\cite{Casteill:2007td} which  
describes the splitting and joining of ${\cal N}=4$ SYM
operators dual to the folded Frolov--Tseytlin
string~\cite{Frolov:2003xy}.

Another interesting and important
line of investigation would be to explicitly relate
non-planar contributions in the
${\cal N}=6$ superconformal Chern--Simons--matter theory to observables in the
dual type IIA string theory.
\vspace*{0.5cm}

\noindent
{\bf Acknowledgments:}
We thank G.\ Grignani, T.\ Harmark, S.\ Hirano and A.\  Wereszczinsky for useful discussions. 
CK and KZ were supported by FNU through grant number 272-06-0434. MO acknowledges FNU
for financial support through grant number 272-08-0050. 
\appendix

\newpage
\section{Derivation of the non-planar dilatation generator \label{derivation2}}

Here we derive explicitly the full two-loop dilatation generator in
the $SU(2)\times SU(2)$ sector using the method of effective vertices
explained in section~\ref{derivation1}.
As already mentioned
the scalar D-terms give rise to the following effective vertex
\beq \label{VD2}
\begin{split}
\left(V_D^{bos}\right)^{eff}= \gamma \, :\,\Tr &\left[
\left(Z^A Z_A^\dg + W^{\dg A} W_A\right)
\left(Z^BZ_B^\dg\!-\!W^{\dg B}W_B\right)
\left(Z^C Z_C^\dg\!-\!W^{\dg C}W_C\right)\right. \\
&  +
\left(Z_A^{\dg} Z^A + W_A W^{\dg A}\right)
\left(Z_B^\dg Z^B \!-\!W_B W^{\dg B}\right)
\left(Z_C^\dg Z^C \!-\! W_C W^{\dg C}\right) \\
&  \!-\! 2 Z_A^\dg \left(Z^BZ_B^\dg\!-\!W^{\dg B}W_B \right) Z^A
\left(Z_C^\dg Z^C \!-\! W_C W^{\dg C}\right)  \\
& \left. \!-\!2 W^{\dg A}\left(Z_B^\dg Z^B \!-\!W_B W^{\dg B}\right) W_A
\left(Z^C Z_C^\dg\!-\!W^{\dg C}W_C\right) \right] : 
\end{split}
\eeq
where $:\,\,:$ means that self-contractions should be omitted.
For the subsequent considerations, it is useful to notice that the
following operator gives a vanishing
contribution when applied to operators of the type appearing in
eqn.~\rf{operators}

\beq 
\begin{split}
V = \gamma \bigg\{ \Tr&\left[
\left(Z^A Z_A^\dg + W^{\dg A} W_A\right)
\left(Z^BZ_B^\dg-W^{\dg B}W_B\right)
\left(Z^C Z_C^\dg-W^{\dg C}W_C\right)\right. \\
&\;+
\left(Z_A^{\dg} Z^A + W_A W^{\dg A}\right)
\left(Z_B^\dg Z^B -W_B W^{\dg B}\right)
\left(Z_C^\dg Z^C - W_C W^{\dg C}\right) \\
  &\;- 2 Z_A^\dg \left(Z^BZ_B^\dg-W^{\dg B}W_B \right) Z^A
\left(Z_C^\dg Z^C - W_C W^{\dg C}\right)
 \\
&\;- 2 W^{\dg A}\left(Z_B^\dg Z^B -W_B W^{\dg B}\right) W_A
\left(Z^C Z_C^\dg-W^{\dg C}W_C\right) \Big] \\
&\; -\left[
N\Tr \left( Z_B^\dg Z^B Z_C^\dg Z^C\right)
-N\Tr \left(Z^B Z_B^\dg Z^C Z_C^\dg\right) \right.  \\
&\;\quad+N\Tr \left(W^{\dg B} W_B W^{\dg C} W_C\right)
-N\Tr \left(W_B W^{\dg B}W_C W^{\dg C}\right)
 \\
&\;\quad +2 N\Tr \left(Z^B Z_B^\dg W^{\dg C} W_C \right)
+2N \Tr \left( W_B W^{\dg B} Z_C^\dg Z^C\right)  \\
&\;\quad+2 \Tr \left(Z^B Z_B^\dg\right)\Tr \left(Z^C Z_C^\dg\right)
+2 \Tr \left(W^{\dg B} W_B\right)\Tr \left(W^{\dg C} W_C\right)  \\
&\;\quad   -2 \Tr \left(Z^B Z_C^\dg \right) \Tr \left(Z^C Z_B^\dg \right)
-2 \Tr \left(W^{\dg B} W_C\right)\Tr \left(W^{\dg C} W_B\right)\\
&\;\quad\left. -4 \Tr \left(Z^B W_C\right) \Tr \left(Z_B^\dg W^{\dg C}\right)
\right] \bigg\}.
\end{split}
\eeq
This can be seen as follows. If we contract $Z_C^\dg$ and $W_C^\dg$ in
the factors $\left(Z^C Z_C^\dg-W^{\dg C}W_C\right)$ in the first four
lines with $W$'s and
$Z$'s inside the operator $\CO$ we get zero. If we contract the same
$Z_C^\dg$ and $W_C^\dg$ with fields inside the vertex itself we get
minus the remaining lines. Notice that there is no normal ordering in the
vertex $V$.

We can rewrite the above effective vertex~\rf{VD2} in the following way
\beq
\begin{split}
\left(V_D^{bos}\right)^{eff}= \gamma \Big\{ &\Tr \left[
\left(Z^A Z_A^\dg + W^{\dg A} W_A\right)
\left(Z^BZ_B^\dg-W^{\dg B}W_B\right)
\left(Z^C Z_C^\dg-W^{\dg C}W_C\right) \right. \\
 &\quad+
\left(Z_A^{\dg} Z^A + W_A W^{\dg A}\right)
\left(Z_B^\dg Z^B -W_B W^{\dg B}\right)
\left(Z_C^\dg Z^C - W_C W^{\dg C}\right)  \\
 &\quad- 2 Z_A^\dg \left(Z^BZ_B^\dg-W^{\dg B}W_B \right) Z^A
\left(Z_C^\dg Z^C - W_C W^{\dg C}\right) \\
 &\quad-2 W^{\dg A}\left(Z_B^\dg Z^B -W_B W^{\dg B}\right) W_A
\left(Z^C Z_C^\dg-W^{\dg C}W_C\right) \Big]   \\
& - : \left[ 3N \Tr \left(Z^B Z_B^\dg Z^C Z_C^\dg\right)
+3N \Tr \left( Z_B^\dg Z^B Z_C^\dg Z^C\right)
\right.  \\
&\quad\;  + 3N \Tr \left(W_B W^{\dg B}W_C W^{\dg C}\right)
+3N \Tr \left(W^{\dg B} W_B W^{\dg C} W_C\right) \\
&\quad\; -2N \Tr \left(Z^B Z_B^\dg W^{\dg C} W_C \right)
-2N \Tr \left( W_B W^{\dg B} Z_C^\dg Z^C\right)  \\
&\quad\;-2 \Tr \left(Z^B Z_B^\dg\right)\Tr \left(Z^C Z_C^\dg\right)
-2 \Tr \left(W^{\dg B} W_B\right)\Tr \left(W^{\dg C} W_C\right)  \\
&\quad\;+12 \Tr \left(Z^B Z_B^\dg\right)\Tr \left(W^{\dg C} W_C\right) \\
&\quad\;-4 \Tr \left(Z^B Z_C^\dg \right) \Tr \left(Z^C Z_B^\dg \right)
-4 \Tr \left(W^{\dg B} W_C\right)\Tr \left(W^{\dg C} W_B\right)\\
&\quad\; \left. -8 \Tr \left(Z^B W_C\right) \Tr \left(Z_B^\dg W^{\dg C}\right)
\right] : \\
& - : \left[18(N^2-1) \Tr\left( Z^C Z_C^\dg\right)+ 18(N^2-1)\Tr \left(W^{\dg C} W_C\right)
\right] : \\
& -24N^2(N^2-1)\Big\}.
\end{split}
\eeq
To this effective vertex we must add the effective vertices corresponding
to the gluon exchange (Fig.~1b), fermion exchange (Fig.~1c) and scalar
self-interactions. 
What we will get if the ``usual'' cancellation takes place is the
vertex $V$. We can rewrite the above vertex without normal ordering
as follows
\bea
\left(V_D^{bos}\right)^{eff}&=&
\mbox{sextic terms} + \mbox{quartic terms} \nonumber \\
&& + 18(N^2-1) \left\{
\Tr\left( Z^C Z_C^\dg\right)+ \Tr \left(W^{\dg C} W_C\right)
\right\}  \nonumber \\
& &-24N^2(N^2-1).
\eea

Let us continue with the fermion exchange, cf.~Fig~1c. It is easy to see that the  term
$V_F^{ferm}$ does not contribute to the anomalous dimension of operators of
the type~\rf{operators}: A diagram like the one in Fig.~1c
requires two fermionic vertices with respectively a daggered and
an undaggered scalar field. Such vertices do not appear in $V_F^{ferm}$.
Furthermore, the first line in $V_D^{ferm}$ can be shown not to give any
contribution.
What remains is an effective vertex which looks like
\beq
\begin{split}
(V^{ferm})^{eff}&= \alpha : \Big\{
N \Tr \left(Z^B Z_B^\dg Z^C Z_C^\dg\right)
+N \Tr \left( Z_B^\dg Z^B Z_C^\dg Z^C\right) \\
&\qquad\quad + N \Tr \left(W_B W^{\dg B}W_C W^{\dg C}\right)
+N \Tr \left(W^{\dg B} W_B W^{\dg C} W_C\right)  \\
&\qquad\quad+4 \Tr \left(Z^B Z_B^\dg\right)\Tr \left(W^{\dg C} W_C\right) \\
&\qquad\quad -2 \Tr \left(Z^B Z_C^\dg \right) \Tr \left(Z^C Z_B^\dg \right)
-2 \Tr \left(W^{\dg B} W_C\right)\Tr \left(W^{\dg C} W_B\right) \\
&\qquad\quad \left. -4 \Tr \left(Z^B W_C\right) \Tr \left(Z_B^\dg W^{\dg C}\right)
\right\} : \\
&= \alpha  \left\{ \mbox{ quartic terms} \right. \\
&\quad\quad - 16(N^2-1) \left[
\Tr\left( Z^C Z_C^\dg\right)+ \Tr \left(W^{\dg C} W_C\right)
\right]  \\
&\quad\quad +32N^2(N^2-1) \big\},
\end{split}
\eeq
where $\alpha$ is a coefficient which is to be determined by Feynman
diagram computations and where {\it quartic terms} means the quartic
terms from before without normal ordering.

Gluon exchange, cf.~Fig~1b gives another contribution
to the anomalous dimension
of the operators in question. The associated effective vertex reads
\beq
\begin{split}
(V^{gluon})^{eff}&=\beta : \left\{ N \Tr \left(Z^B Z_B^\dg Z^C Z_C^\dg\right)
+N \Tr \left( Z_B^\dg Z^B Z_C^\dg Z^C\right)
\right.  \\
&\quad\qquad + N \Tr \left(W_B W^{\dg B}W_C W^{\dg C}\right)
+N \Tr \left(W^{\dg B} W_B W^{\dg C} W_C\right) \\
&\quad\qquad +2N \Tr \left(Z^B Z_B^\dg W_C^\dg W^C\right)
+ 2N \Tr \left(Z_B^\dg Z^B W^C W_C^\dg\right) \\
&\quad\qquad +2\Tr \left(Z^B Z_B^\dg\right)\Tr \left(Z^C Z_C^\dg\right)
+ 2\Tr \left(W^{\dg B} W_B\right)\Tr \left(W^{\dg C} W_C\right)\\
 &\quad\qquad+4 \Tr \left(Z^B Z_B^\dg\right)\Tr \left(W^{\dg C} W_C\right)  \\
&\quad\qquad -4 \Tr \left(Z^B Z_C^\dg \right) \Tr \left(Z^C Z_B^\dg \right)
-4 \Tr \left(W^{\dg B} W_C\right)\Tr \left(W^{\dg C} W_B\right) \\
&\quad\qquad \left. -8 \Tr \left(Z^B W_C\right) \Tr \left(Z_B^\dg W^{\dg C}\right)
\right\} : \\
&=
\beta  \left\{ \mbox{ quartic terms} \right.\\
&\qquad - 28(N^2-1) \left[
\Tr\left( Z^C Z_C^\dg\right)+ \Tr \left(W^{\dg C} W_C\right)
\right]  \\
&\qquad\left. +56N^2(N^2-1) \right\},
\end{split}
\eeq
where $\beta$ is a coefficient which likewise is to be determined by Feynman
diagram computations.

Noticing that the scalar self-interactions can never give a contribution to
the effective vertex which mixes different indices inside the same trace
we find that in order that the expected cancellation takes place we need that
\beq
\alpha=\gamma-2\beta.
\eeq
Inserting this we find
\bea
\lefteqn{(V_D^{bos})^{eff}+(V^{ferm})^{eff}+(V^{gluon})^{eff}-V =}
 \nonumber   \\
&&
(\beta+ 3\gamma)N \left\{\Tr \left(Z^B Z_B^\dg
\left( W^{\dg C} W_C- Z^C Z_C^\dg\right)\right)+
\Tr \left(W_B W^{\dg B}\left(Z_C^\dg Z^C -W_C W^{\dg C}\right )\right)\right\}
\nonumber \\
& & +(\beta+\gamma)N \left\{
\Tr \left( Z_B^\dg Z^B \left( W_C W^{\dg C} -Z_C^\dg Z^C\right)\right) +
\Tr \left(W^{\dg B} W_B \left (Z^C Z_C^\dg -W^{\dg C} W_C\right)\right)
\right\}
\nonumber \\
&& +(2\beta +4\gamma)\Tr \left(Z^B Z_B^\dg - W^{\dg B} W_B\right)
\Tr \left(Z^C Z_C^\dg - W^{\dg C} W_C\right)\nonumber \\
& & +(4 \beta+2\gamma) (N^2-1) \left \{\Tr \left(Z^C Z_C^\dg\right)
+ \Tr \left(W^{\dg C} W_C\right) \right\} +(8\gamma-8\beta) N^2(N^2-1).
\eea
As already exploited, terms containing factors of the type
$\left(Z_C^\dg Z^C -W_C W^{\dg C}\right )$ only give a non-vanishing
contribution when $Z_C^\dg$ and $W^{\dg C}$ are contracted with fields
inside the vertex itself. Therefore, we have
\bea
\lefteqn{(V_D^{bos})^{eff}+(V^{ferm})^{eff}+(V^{gluon})^{eff}-V }\nonumber \\
&=&
(2\beta-2\gamma)(N^2-1) :\left\{\Tr\left( Z_C^\dg Z^C\right)+
\Tr\left(W_C W^{\dg C}\right)\right\}:
\label{Veff-V}
\eea
This exactly has the form expected for scalar self-interactions. Now we have
to determine the coefficients and check that everything fits. From
reference~\cite{Bak:2008cp} we can read off the values of $\gamma$ and
$\beta$. They are
\beq
\gamma=\frac{1}{4}\frac{\lambda^2}{N^2},
\hspace{0.7cm} \beta= -\frac{1}{8}\frac{\lambda^2}{N^2}.
\eeq
This means that we need that
\beq
\alpha= \frac{1}{2}\frac{\lambda^2}{N^2},
\eeq
which can easily be verified using reference~\cite{Bak:2008cp}.
Finally we find
for the pre-factor in eqn.~\rf{Veff-V}
\beq
(2\beta-2\gamma)(N^2-1)= 
-\frac{3}{4}\lambda^2\left(1-\frac1{N^2}\right).
\eeq
This is exactly equal to minus the pre-factor of the scalar self-energies.
The planar part can again be read off directly from \cite{Bak:2008cp}, while
to verify the term subleading in $N^2$ we performed a closer analysis 
of the non--planar versions of the self--energy diagrams. Thus, we have shown
that the full one-loop dilatation generator in the $SU(2)\times SU(2)$ sector
is indeed given only by the $F$-terms in the bosonic potential.

\section{Subleading contributions for BMN states}

\subsection{Operators with only one type of excitation \label{Boperators}}
Below we present the contributions to $ D \CB_p^{J_0,J_1,\ldots,J_k}$ which
are of order $\frac{1}{N^2}$, cf.\ eqn.\rf{Dexpansion}. As mentioned in the
main text none of these terms survive in the BMN limit. As the terms
are multiplied by $\frac{\lambda^2}{N^2}$ they need to be of the
order $J^2$ to contribute in the limit. However, the maximum order of
any term is $J$. All terms involve operators in a combination which
turns into a first derivative in the BMN limit and which is thus of order
$\frac{1}{J}$. At the same time any term can at maximum contain two
sums (arising via the second and third contraction)
each giving a factor of $J$.

\begin{eqnarray}
\lefteqn{D_{++}\circ\CB_{p}^{J_0,J_1,\ldots,J_k}= (-2) \,\,
 \times\label{H++B}}\\
 &&
\hspace*{-.5cm}
\left[\sum_{J_{k+2}=1}^{p-J_{k+1}-2}\sum_{J_{k+1}=1}^{p-2}
\left(\CB_{p-J_{k+1}-J_{k+2}-2}^{J_0-J_{k+1}-J_{k+2}-1,J_1,\ldots,J_k,J_{k+1},J_{k+2}}
-\CB_{p-J_{k+1}-J_{k+2}-1}^{J_0-J_{k+1}-J_{k+2}-1,J_1,\ldots,J_k,J_{k+1},J_{k+2}}\right)\right.\nonumber \\
&&
\hspace*{-.8cm}
+
\left.\sum_{J_{k+2}=1}^{J_0-p-J_{k+1}-2}\sum_{J_{k+1}=1}^{J_0-p-2}\left(\CB_{p}^{J_0-J_{k+1}-J_{k+2}-1,J_1,\ldots,J_k,J_{k+1},J_{k+2}}
-\CB_{p+1}^{J_0-J_{k+1}-J_{k+2}-1,J_1,\ldots,J_k,J_{k+1},J_{k+2}}\right)\right],
~~~~~  \nonumber
\end{eqnarray}
\beq\label{H--B}
\begin{split}
D_{--}\circ\CB_{p}^{J_0,J_1,\ldots,J_k}=&\\
- 2\bigg[\sum_{i=1}^{k}J_i\sum_{j\neq i}^{k}J_j& 
\left(\CB_{J_i+J_j+p-1}^{J_0+J_i+J_j,J_1,\ldots,
\makebox[0pt]{\,\,\,\,$\times$}J_i,
\ldots,\makebox[0pt]{\,\,\,\,$\times$}J_j, \ldots,J_k}
-\CB_{J_i+J_j+p}^{J_0+J_i+J_j,J_1,\ldots,
\makebox[0pt]{\,\,\,\,$\times$}J_i,
\ldots,\makebox[0pt]{\,\,\,\,$\times$}J_j,
\ldots,J_k}\right.\\
&-\left.\left.\CB_{p}^{J_0+J_i+J_j,J_1,\ldots,
\makebox[0pt]{\,\,\,\,$\times$}J_i,
\ldots,\makebox[0pt]{\,\,\,\,$\times$}J_j, \ldots,J_k}
+\CB_{p+1}^{J_0+J_i+J_j,J_1,\ldots,
\makebox[0pt]{\,\,\,\,$\times$}J_i,
\ldots,\makebox[0pt]{\,\,\,\,$\times$}J_j, \ldots,J_k}\right)\right],
\end{split}
\eeq
\begin{eqnarray}
\lefteqn{D_{00}\circ\CB_{p}^{J_0,J_1,\ldots,J_k}= }\\
&&-\left[2\sum_{p=0}^{J_0-1}\left(
\CB_{p}^{J_0,J_1,\ldots,J_k}-\CB_{p+1}^{J_0,J_1,\ldots,J_k}\right)+
p(p+1)\left(\CB_{p-1}^{J_0,J_1,\ldots,J_k}-\CB_{p}^{J_0,J_1,\ldots,J_k}\right)
\right.\nonumber \\
&&+\left.
({J}_0-p)({J}_0-p+1)\left(\CB_{p+1}^{J_0,J_1,\ldots,J_k}-\CB_{p}^{J_0,J_1,\ldots,J_k}\right)\right.\nonumber \\
&&+\left.\sum_{l=0}^{p-1}\left(
\CB_{p-l-1}^{J_0,J_1,\ldots,J_k}-\CB_{p-l}^{J_0,J_1,\ldots,J_k}\right)+
\sum_{l=0}^{J_0-p-1}\left(
\CB_{p+l+1}^{J_0,J_1,\ldots,J_k}-
\CB_{p+l}^{J_0,J_1,\ldots,J_k}\right)\right.\cr
&&+\sum_{s=0}^{J_0-l-1}\left(\sum_{l=0}^{p-1}+\sum_{l=0}^{J_0-p-1}\right)\left(
\CB_{l+s+1}^{J_0,J_1,\ldots,J_k}-\CB_{l+s}^{J_0,J_1,\ldots,J_k}\right) \cr
&&
+\sum_{s=0}^{J_0-l-2}\left(\sum_{l=0}^{p-1}+\sum_{l=0}^{J_0-p-1}\right)\left(
\CB_{s}^{J_0,J_1,\ldots,J_k}-\CB_{s+1}^{J_0,J_1,\ldots,J_k}\right)\cr
&&+\left.\sum_{i=1}^k
J_i\left(\sum_{l=0}^{J_0-p-1}+\sum_{l=0}^{J_0-p-2}\right)
\left(\CB_{p+1}^{J_0+J_i-l-1,J_1,\ldots,
\makebox[0pt]{\,\,\,\,$\times$}J_i, \ldots,J_k,l+1}
-\CB_{p}^{J_0+J_i-l-1,J_1,\ldots,
\makebox[0pt]{\,\,\,\,$\times$}J_i, \ldots,J_k,l+1}\right)\right.\cr
&&+\left.\sum_{i=1}^k J_i\left(\CB_{p+1}^{p+J_i,J_1,\ldots,
\makebox[0pt]{\,\,\,\,$\times$}J_i, \ldots,J_k,J_0-p}
-\CB_{p}^{p+J_i,J_1,\ldots, \makebox[0pt]{\,\,\,\,$\times$}J_i,
\ldots,J_k,J_0-p}\right)\right.\cr &&+\left.\sum_{i=1}^k
J_i\left(\CB_{J_0-p+1}^{J_0-p+J_i,J_1,\ldots,
\makebox[0pt]{\,\,\,\,$\times$}J_i, \ldots,J_k,p}
-\CB_{J_0-p}^{J_0-p+J_i,J_1,\ldots,
\makebox[0pt]{\,\,\,\,$\times$}J_i, \ldots,J_k,p}\right) \right.\cr
&&+\left.2\sum_{i=1}^k
J_i\sum_{l=0}^{J_0+J_i-p-2}\left(\CB_{p+1}^{J_0+J_i-l-1,J_1,\ldots,
\makebox[0pt]{\,\,\,\,$\times$}J_i, \ldots,J_k,l+1}
-\CB_{p}^{J_0+J_i-l-1,J_1,\ldots,
\makebox[0pt]{\,\,\,\,$\times$}J_i, \ldots,J_k,l+1}\right)
\right.\cr &&+\left.\sum_{i=1}^k
J_i\left(\sum_{l=0}^{p-1}+\sum_{l=0}^{p-2}\right)
\left(\CB_{p+J_i-l-2}^{J_0+J_i-l-1,J_1,\ldots,
\makebox[0pt]{\,\,\,\,$\times$}J_i, \ldots,J_k,l+1}
-\CB_{p+J_i-l-1}^{J_0+J_i-l-1,J_1,\ldots,
\makebox[0pt]{\,\,\,\,$\times$}J_i, \ldots,J_k,l+1}\right)
\right.\cr &&+\left.2\sum_{i=1}^k
J_i\sum_{l=0}^{p+J_i-2}\left(\CB_{p+J_i-l-2}^{J_0+J_i-l-1,J_1,\ldots,
\makebox[0pt]{\,\,\,\,$\times$}J_i, \ldots,J_k,l+1}
-\CB_{p+J_i-l-1}^{J_0+J_i-l-1,J_1,\ldots,
\makebox[0pt]{\,\,\,\,$\times$}J_i, \ldots,J_k,l+1}\right) \right].\nonumber
\end{eqnarray}
\subsection{Operators with two different types of excitations
\label{Aoperators}}
Below we present the $\frac{1}{N^2}$-contributions to
$ D \CA_p^{J_0,J_1,\ldots,J_k}$, cf. eqn.\rf{Dexpansion}.
As in the case of the $\CB$-operators
and for the same reason none
of these terms survive in the BMN limit, cf.\ Appendix~\rf{Boperators}.
\begin{eqnarray}
\hspace*{-1.0cm}
\lefteqn{D_{++}\circ\CA_{p}^{J_0,J_1,\ldots,J_k}=}\cr
 &&-2
\left[\sum_{J_{k+2}=1}^{p-J_{k+1}-2}\sum_{J_{k+1}=1}^{p-2}
\left(\CA_{p-J_{k+1}-J_{k+2}-2}^{J_0-J_{k+1}-J_{k+2}-1,J_1,\ldots,J_k,J_{k+1},J_{k+2}}
\!-\!\CA_{p-J_{k+1}-J_{k+2}-1}^{J_0-J_{k+1}-J_{k+2}-1,J_1,\ldots,J_k,J_{k+1},J_{k+2}}\right)\right.\cr
&&-\left.
\sum_{J_{k+2}=1}^{J_0-p-J_{k+1}-2}\sum_{J_{k+1}=1}^{J_0-p-2}\left(\CA_{p}^{J_0-J_{k+1}-J_{k+2}-1,J_1,\ldots,J_k,J_{k+1},J_{k+2}}
\!-\!\CA_{p+1}^{J_0-J_{k+1}-J_{k+2}-1,J_1,\ldots,J_k,J_{k+1},J_{k+2}}\right)
\right.\cr &&+\left.\sum_{J_{k+1}=1}^{p-1}
\left(\CA_{0}^{p-J_{k+1},J_1,\ldots,J_k,J_{k+1},J_0-p}
-\CA_{p-J_{k+1}}^{p-J_{k+1},J_1,\ldots,J_k,J_{k+1},J_0-p}\right)\right.\cr
&&+\left.\sum_{J_{k+1}=1}^{J_0-p-1}\left(\CA_{J_0-p-J_{k+1}}^{J_0-p-J_{k+1},J_1,\ldots,J_k,J_{k+1},p}
-\CA_{0}^{J_0-p-J_{k+1},J_1,\ldots,J_k,J_{k+1},p}\right)
\right],~~~~~ \label{H++} \\
&& \nonumber
\end{eqnarray}
\beq \label{H--}
\begin{split}
D_{--}\circ\CA_{p}^{J_0,J_1,\ldots,J_k}=&\cr
 -2
\sum_{i=1}^{k}J_i\sum_{j\neq
i}^{k}J_j&\left[\CA_{J_i+J_j+p-1}^{J_0+J_i+J_j,J_1,\ldots,
\makebox[0pt]{\,\,\,\,$\times$}J_i,
\ldots,\makebox[0pt]{\,\,\,\,$\times$}J_j, \ldots,J_k}
-\CA_{J_i+J_j+p}^{J_0+J_i+J_j,J_1,\ldots,
\makebox[0pt]{\,\,\,\,$\times$}J_i,
\ldots,\makebox[0pt]{\,\,\,\,$\times$}J_j, \ldots,J_k}\right.\cr
&-\left.\CA_{p}^{J_0+J_i+J_j,J_1,\ldots,
\makebox[0pt]{\,\,\,\,$\times$}J_i,
\ldots,\makebox[0pt]{\,\,\,\,$\times$}J_j, \ldots,J_k}
+\CA_{p+1}^{J_0+J_i+J_j,J_1,\ldots,
\makebox[0pt]{\,\,\,\,$\times$}J_i,
\ldots,\makebox[0pt]{\,\,\,\,$\times$}J_j, \ldots,J_k}\right],
\end{split}
\eeq
\begin{eqnarray}
\lefteqn{D_{00}\circ\CA_{p}^{J_0,J_1,\ldots,J_k}= -\left[p(p-1)\left(\CA_{p-1}^{J_0,J_1,\ldots,J_k}-\CA_{p}^{J_0,J_1,\ldots,J_k}\right)
\right.}
\cr
&&+\left. ({J}_0-p)({J}_0-p-1)\left(
\CA_{p+1}^{J_0,J_1,\ldots,J_k}-\CA_{p}^{J_0,J_1,\ldots,J_k}\right)
\right.\cr
&&+\left.2p\left(\CA_{J_0}^{J_0,J_1,\ldots,J_k}-\CA_{J_0-1}^{J_0,J_1,\ldots,J_k}\right)
+
2(J_0-p)\left(\CA_{J_0-p-1}^{J_0,J_1,\ldots,J_k}-\CA_{J_0-p}^{J_0,J_1,\ldots,J_k}\right)
\right.\cr &&+\left.
2\left(\CA_{J_0}^{J_0,J_1,\ldots,J_k}-\CA_{0}^{J_0,J_1,\ldots,J_k}\right)
(p\,\delta_{p\neq 0}+(J_0-p)\delta_{p\neq J_0}) \right.\cr &&+\left.
\sum_{l=0}^{p-1}\sum_{s=0}^{J_0-l-2}\left(\CA_{J_0-s-1}^{J_0,J_1,\ldots,J_k}-\CA_{J_0-s-2}^{J_0,J_1,\ldots,J_k}\right)
+
\sum_{l=0}^{p-1}\sum_{s=0}^{J_0-p-l-1}\left(\CA_{p-l+s}^{J_0,J_1,\ldots,J_k}-\CA_{p-l+s-1}^{J_0,J_1,\ldots,J_k}\right)
\right.\cr &&+\left.
\sum_{l=0}^{J_0-p-1}\sum_{s=0}^{J_0-l-2}\left(\CA_{s}^{J_0,J_1,\ldots,J_k}-\CA_{s+1}^{J_0,J_1,\ldots,J_k}+
\CA_{J_0-l-s-2}^{J_0,J_1,\ldots,J_k}-\CA_{J_0-l-s-1}^{J_0,J_1,\ldots,J_k}\right)
\right.\cr &&+\left. 2\sum_{i=1}^k J_i\sum_{l=0}^{p-2}
\left(\CA_{p+J_i-l-2}^{J_0+J_i-l-1,J_1,\ldots,
\makebox[0pt]{\,\,\,\,$\times$}J_i, \ldots,J_k,l+1}
-\CA_{p+J_i-l-1}^{J_0+J_i-l-1,J_1,\ldots,
\makebox[0pt]{\,\,\,\,$\times$}J_i, \ldots,J_k,l+1}\right)
\right.\cr &&+\left. 2\sum_{i=1}^k J_i\sum_{l=0}^{J_0-p-2}
\left(\CA_{p+1}^{J_0+J_i-l-1,J_1,\ldots,
\makebox[0pt]{\,\,\,\,$\times$}J_i, \ldots,J_k,l+1}
-\CA_{p}^{J_0+J_i-l-1,J_1,\ldots,
\makebox[0pt]{\,\,\,\,$\times$}J_i, \ldots,J_k,l+1}\right)
\right.\cr &&+\left. 2\sum_{i=1}^k J_i\sum_{l=0}^{J_0-p+J_i-2}
\left(\CA_{p+1}^{J_0+J_i-l-1,J_1,\ldots,
\makebox[0pt]{\,\,\,\,$\times$}J_i, \ldots,J_k,l+1}
-\CA_{p}^{J_0+J_i-l-1,J_1,\ldots,
\makebox[0pt]{\,\,\,\,$\times$}J_i, \ldots,J_k,l+1}\right)
\right.\cr &&+\left. 2\sum_{i=1}^k J_i\sum_{l=0}^{p+J_i-2}
\left(\CA_{p+J_i-l-2}^{J_0+J_i-l-1,J_1,\ldots,
\makebox[0pt]{\,\,\,\,$\times$}J_i, \ldots,J_k,l+1}
-\CA_{p+J_i-l-1}^{J_0+J_i-l-1,J_1,\ldots,
\makebox[0pt]{\,\,\,\,$\times$}J_i, \ldots,J_k,l+1}\right)
\right.\cr &&+\left.2\sum_{i=1}^k
J_i\left(\CA_{0}^{p+J_i,J_1,\ldots,
\makebox[0pt]{\,\,\,\,$\times$}J_i, \ldots,J_k,J_0-p}
-\CA_{p+J_i}^{p+J_i,J_1,\ldots, \makebox[0pt]{\,\,\,\,$\times$}J_i,
\ldots,J_k,J_0-p}\right.\right.\cr
&&+\left.\left.\CA_{J_0-p+J_i}^{J_0-p+J_i,J_1,\ldots,
\makebox[0pt]{\,\,\,\,$\times$}J_i, \ldots,J_k,p}
-\CA_{0}^{J_0-p+J_i,J_1,\ldots, \makebox[0pt]{\,\,\,\,$\times$}J_i,
\ldots,J_k,p} \right)  \right]. \label{H+-}
\end{eqnarray}

\providecommand{\href}[2]{#2}\begingroup\raggedright\endgroup

\end{document}